\documentclass[11pt]{article}
\pdfoutput=1
\usepackage[margin=1.0in]{geometry}
\usepackage{amsmath,amssymb,graphicx}
\usepackage{dsfont}
\usepackage{xcolor}
\definecolor{Mahogany}{rgb}{0.62,0.24,0.15}
\definecolor{DarkRed}{rgb}{0.6,0,0}
\definecolor{DarkGreen}{rgb}{0,0.6,0}
\definecolor{DarkBlue}{rgb}{0,0,0.6}
\definecolor{gray}{RGB}{128,128,128}
\usepackage[colorlinks=true,linktocpage=true,linkcolor=DarkBlue,citecolor=DarkGreen,urlcolor=DarkGreen]{hyperref}
\usepackage{slashed}
\usepackage{verbatim}
\usepackage{cite}

\usepackage{setspace}
\usepackage[T1]{fontenc}

\usepackage[utf8]{inputenc}

\usepackage{upgreek}
\usepackage{bbm} 

\usepackage[font=small,labelfont=bf]{caption}
\let\OLDthebibliography\thebibliography
\renewcommand\thebibliography[1]{
    \small
    \OLDthebibliography{#1}
    \setlength{\parskip}{0.0pt plus 0.5pt}
    \setlength{\itemsep}{3.5pt plus 2.0pt minus 1.0pt}
}

\usepackage{multirow}

\newcommand{\lab}[1]{{\mathrm{#1}}}

\def\e{{\lab{e}}}

\DeclareMathOperator{\tr}{tr}
\DeclareMathOperator{\Tr}{Tr}
\DeclareMathOperator{\diag}{diag}

\renewcommand{\Im}{\mathrm{Im}}

\def\dd{\text{d}}
\newcommand{\hc}{{\textrm{h.c.}}}
\newcommand{\thetabar}{\overline{\theta}}

\newcommand{\minus}{{\scalebox {0.75}[1.0]{$-$}}}

\def\LCP{\Lambda_{\textsc{CP}}}
\def\Left{\Lambda_{\textsc{eft}}}

\newcommand{\tlambda}{\tilde{\lambda}}
\newcommand{\tkappa}{\tilde{\kappa}}

\newcommand{\be}{\begin{equation}}
\newcommand{\ee}{\end{equation}}
\newcommand{\bp}{\begin{pmatrix}}
\newcommand{\ep}{\end{pmatrix}}
\newcommand{\ba}{\begin{aligned}}
\newcommand{\ea}{\end{aligned}}

\newcommand{\bealg}{\begin{equation}\begin{aligned}} 
\newcommand{\eealg}{\end{aligned}\end{equation}} 

\def\gev{\;\hbox{GeV}}

\title{\bf Strong CP from a Hidden Chiral Condensate}

\author{
Csaba Cs\'aki$^{1}$, Samuel Homiller$^{1}$ and Taewook Youn$^{1,2}$
\\[0.5em]
{\small \color{gray} \texttt{csaki@cornell.edu},~
\texttt{shomiller@cornell.edu},~
\texttt{taewook.youn@cornell.edu}}\\[0.5em]
{\small $^{1}$Laboratory for Elementary Particle Physics, Cornell University, Ithaca, NY 14853}\\[0.25em] 
{\small $^{2}$School of Physics, Korea Institute for Advanced Study, Seoul 02455, Republic of Korea}\\[0.25em]
}

\begin{document}

\maketitle

\vspace*{-2em}

\begin{abstract}
Models which solve the strong CP problem by employing discrete spacetime symmetries generically suffer fine-tuning and quality problems. We demonstrate that these issues are greatly ameliorated when the only source of spontaneous CP breaking is from the chiral condensate of a strongly coupled hidden sector. This is shown explicitly in a model with the SM extended by a vector-like quark family and a complex scalar portal to QCD-like dark sector with $N_f$ families of dark fermions that confines at a high scale. The dark pions of the hidden sector are natural dark matter candidates, with the correct relic abundance obtained via freeze-in. These ``confining'' Nelson--Barr solutions connect phenomenological questions regarding the strong CP problem to recent developments in the understanding of confining gauge theories, and present ample room for further model building.
\end{abstract}

\vskip 0.5cm

\section{Introduction}

Studying the nature of discrete spacetime symmetries has played a pivotal role in the development of the Standard Model (SM) and our understanding of quantum field theory.
The discovery of parity violation in weak decays presaged the structure of the standard electroweak theory, while the discovery of CP violation in the weak interactions cemented our understanding of the SM particle content and opened new avenues for testing its internal consistency.

There are two sources of CP violation in the renormalizable part of the Standard Model: the Jarlskog invariant~\cite{Jarlskog:1985ht} and the strong CP phase,
\begin{equation}
\theta_{\textrm{weak}} \propto \arg \det\big( Y_u Y_d - Y_d Y_u \big) \,,
\qquad 
\thetabar = \theta - \arg \det\big(Y_u Y_d\big) \, .
\end{equation}
Here, $Y_u$ and $Y_d$ are the Yukawa coupling matrices of up- and down-type quarks in a Hermitian flavor basis and $\theta$ is the coefficient of the topological term in the QCD Lagrangian:
\begin{equation}
\mathcal{L} \supset \frac{\theta}{16\pi^2} \tr G_{\mu\nu} \widetilde{G}^{\mu\nu} \, .
\end{equation}
The strong CP problem is the puzzle that, while both of these parameters are on equal footing with regards to the symmetries of the Standard Model, $\theta_{\textrm{weak}}$ is $\mathcal{O}(1)$ while experimental constraints on the neutron EDM require $\thetabar \lesssim 10^{-10}$.  

The only symmetries under which $\thetabar$ transforms non-trivially are the {\em spacetime} symmetries P and CP, under which $\thetabar$ is odd.
This suggests a symmetry-based solution to the strong CP puzzle, where P or CP is promoted to an exact symmetry at high scales, spontaneously broken in our vacuum. 
This is particularly motivated by the existence of UV-complete gravitational theories in which P/CP arises as a discrete symmetry in higher dimensions compactified to 4d~\cite{Strominger:1985it} (see also refs.~\cite{Dine:1992ya, Choi:1992xp}).
While parity-based models are interesting---see refs.~\cite{Chakdar:2013tca, Hall:2018let, Dunsky:2019api, Craig:2020bnv, deVries:2021pzl} for a selection of recent papers---they require a much more significant extension to the SM field content than CP-based solutions, so we will focus on the latter in what follows.

If CP is assumed to be a symmetry of the UV, the challenge in solving the strong CP problem is as much generating $\theta_{\textrm{weak}}$ to be close to maximal as it is in keeping $\thetabar$ small. 
Crucially, such a pattern would be radiatively stable once generated: the running of $\thetabar$ in the SM arises only at seven loops~\cite{Ellis:1978hq}.\footnote{There are finite corrections at four-loops from ``Cheburashka diagrams'' found by Khriplovich~\cite{Khriplovich:1985jr}, but these are too small to spoil the CP-based solution to the strong CP problem.}

Nelson and Barr demonstrated that an $\mathcal{O}(1)$ CKM phase with $\thetabar = 0$ is natural in a broad class of theories, depending on the additional fermion representations and how they are charged under symmetries beyond the SM~\cite{Nelson:1983zb, Barr:1984qx, Barr:1984fh, Nelson:1984hg}.
While originally developed in the context of more complicated, grand unified theories\footnote{It was recently emphasized in ref.~\cite{Vecchi:2025qie} that grand unified models may also provide an explanation for why the CP-preserving point should be $\thetabar = 0$ rather than $\pi$. See also the discussion therein, and in ref.~\cite{Kuchimanchi:2025pqb}, on the validity of these symmetry-based solutions to the strong CP problem.}, a minimal implementation of this ``Nelson--Barr mechanism'' was developed by Bento, Branco and Parada~\cite{Bento:1991ez} requiring only one additional set of quarks, transforming as a vector-like pair under the SM.
Other routes to preserving $\thetabar = 0$ after spontaneous CP breaking involve enforcing the Hermiticity of the Yukawa matrices with supersymmetry~\cite{Hiller:2001qg, Hiller:2002um} or via twisted split fermions in extra-dimensional models~\cite{Harnik:2004su}, or by imposing additional flavor symmetries in an extended Higgs sector~\cite{Hall:2024qqe, Bonnefoy:2025rvo}.

Common to all of these symmetry-based solutions is a new scale, $\LCP$: the scale at which CP is spontaneously broken.
Studying the interplay between the nature of CP-violation in the strong interactions and the flavor sector thus directly tests fundamental physics at a much higher scale. 
In the simplest versions of these theories, $\LCP$ is the vacuum expectation value of a scalar field, which can generally be complex even if all the parameters of the potential are real~\cite{Haber:2012np}.

On the other hand, the existence of this scale presents a new challenge: in effective field theory, we naturally expect corrections to $\thetabar$ from higher-dimension operators, suppressed by powers of the cutoff, $\Left$~\cite{Dine:2015jga, Perez:2020dbw, Valenti:2021xjp, Asadi:2022vys}. If these corrections arise at dimension-five, they naively contribute $\Delta\thetabar \sim \LCP / \Left$, which is too large for $\LCP \gtrsim 10^8\,\textrm{GeV}$, even if the cutoff is taken to the Planck scale.
This ``quality problem'' is entirely equivalent to the well-known axion quality problem, in which the axion decay constant is the physical scale.
In the Nelson--Barr case, the upper bound on $\LCP$ leads to various model-building challenges to mitigate the presence of cosmological CP-domain walls while allowing for a successful model of inflation and baryogenesis~\cite{McNamara:2022lrw, Asadi:2022vys, Murgui:2025scx}.

Moreover, at scales below $\LCP$, the point $\thetabar = 0$ is no longer radiatively stable. If $\LCP$ is a scalar vacuum expectation value (vev), potential terms mixing the scalar and the SM Higgs---which cannot be forbidden by ordinary internal symmetries---lead to loop corrections to $\thetabar$ (see also the recent discussion in ref.~\cite{Valenti:2021rdu}). As a result, a significant degree of fine-tuning is mandated to preserve $\thetabar \lesssim 10^{-10}$.

One approach to these problems is to introduce additional symmetries which forbid corrections to $\thetabar$ at dimension-five~\cite{Dine:2015jga, Valenti:2021xjp, Asadi:2022vys, Perez:2023zin}. These ``chiral'' models generally require additional field content and new symmetries that are themselves of high-quality. Even then, they must be augmented by e.g., supersymmetry, to avoid issues of fine-tuning in both loop corrections to $\thetabar$ and in corrections to the Higgs hierarchy. Supersymmetric models suffer from numerous model-building challenges elucidated in refs.~\cite{Dine:1993qm, Dine:2015jga}; for instance, the spontaneous CP breaking sector must be sequestered from the SUSY-breaking sector in order to avoid both experimental constraints on new sources of CP-violation (e.g., in $K-\bar{K}$ mixing) and too-large contributions to $\thetabar$. 

\medskip

In this paper, we demonstrate that these issues are resolved automatically when CP is spontaneously broken by the dynamics of a confining hidden sector with $\thetabar = \pi$~\cite{Dashen:1970et,Smilga:1998dh,Gaiotto:2017tne,Gaiotto:2017yup,DiVecchia:2017xpu,Cordova:2019bsd,Cordova:2019jnf,Cordova:2019uob,Creutz:2003xu,Creutz:2006ts,Draper:2018mpj}.
In such a scenario, the CP violation is communicated to the SM only through operators of higher dimension, and the quality and tuning problems are dealt with automatically. 
This is qualitatively similar to the advantages of composite axion models (see e.g., refs.~\cite{Kim:1984pt, Kaplan:1985dv, Choi:1985cb, Randall:1992ut, Cheng:2001ys, Fukuda:2017ylt, DiLuzio:2017tjx, Lillard:2017cwx, Lillard:2018fdt, Contino:2021ayn}).
The possibility in Nelson--Barr models was previously explored in refs.~\cite{Perez:2020dbw}, but the possibility that the only source of CP-violation could be the chiral condensate was considered less desirable because it was thought that the CP-violating phase was $\mathcal{O}(1)$ only in a small region of the parameter space.\footnote{Ref.~\cite{Valenti:2021xjp} also realizes this protection to $\thetabar$ in a strongly coupled scenario, but with the CP-violating phase arising in an effective scalar potential in a more complicated theory, rather than from the dynamics of the QCD-like theory itself.}

Recently, however, there has been a great deal of progress in understanding the CP phase structure of QCD-like theories at $\thetabar=\pi$. This was brought about by perturbing a supersymmetric (SUSY) version of QCD via anomaly mediated supersymmetry breaking (AMSB)~\cite{Murayama:2021xfj, Csaki:2022cyg}. See refs.~\cite{Csaki:2023yas,Csaki:2025atr} for additional discussions of the phenomenological implications from AMSB QCD.
In particular, it was demonstrated in ref.~\cite{Csaki:2024lvk} that for more than one quark flavor, CP symmetry is always spontaneously broken provided the quark masses are equal and non-zero.
This phenomenon of spontaneous CP breaking with non-zero and degenerate quark masses is verified in the large-$N$ chiral Lagrangian framework~\cite{DiVecchia:2017xpu}. Consequently, a broad range of hidden quark masses can give rise to a sizable CP-violating phase in this scenario as long there is an (approximate) flavor symmetry. This flavor symmetry can have ${\cal O} (1)$ breaking terms, as we explain in Appendix~\ref{app:cp}.

This setup brings about another possibility that has not been previously explored: in the broad parameter space of interest, the strongly-coupled hidden sector may contain stable bound states, neutral under the SM gauge group and only weakly coupled to the visible sector. These are natural candidates for dark matter, and indeed, we demonstrate explicitly that in a minimal implementation of this model, the correct relic abundance for the stable dark pions is achieved via the ``freeze-in'' mechanism across a broad range of parameter space where the strong CP problem is solved.

The rest of this paper is structured as follows:
In Section~\ref{sec:model}, we describe a minimal implementation of our model, and derive the conditions under which the strong CP problem is solved.
In Section~\ref{sec:dm_abundance}, we compute the freeze-in abundance for the dark pions, and find parameter space where the DM and strong CP problems are simultaneously resolved.
We conclude with an extended discussion on future directions in Section~\ref{sec:discussion}. 
In Appendix~\ref{app:cp} we work out the details of spontaneous CP-breaking in a supersymmetric QCD-like theory extended with a complex scalar, demonstrating explicitly under which conditions an $\mathcal{O}(1)$ CP-violating phase arises.

\section{CP Violation from a Confining Hidden Sector}
\label{sec:model}

In this section, we introduce a model in which the strong CP puzzle is resolved by the Nelson--Barr mechanism, with CP violation arising from the dynamics of a hidden sector.
Since CP is a symmetry, the $\thetabar$ parameter associated with QCD is zero in the UV. The important point will be to generate an $\mathcal{O}(1)$ CKM phase in the SM quark mixing matrix without reintroducing $\thetabar$ in the IR. 

We first outline a minimal model implementing this setup, describe how the CP-violating phase in the hidden sector dynamics is communicated to the visible sector, and derive the conditions under which the CKM phase can be $\mathcal{O}(1)$. We then tabulate the one-loop corrections and the effects of non-renormalizable operators. As we will see, because the CP-violation arises only in an operator of dimension three, these unwanted contributions are much more easily suppressed than in typical Nelson--Barr models.

\subsection{A Minimal Model}
\label{subsec:setup}

We now describe a minimal implementation of our setup. 
The model is similar to the one considered in ref.~\cite{Perez:2020dbw}, with slightly different assumptions.
We suppose the existence of a hidden sector with gauge group $\lab{SU}(N)$ that contains $N_f \geq 1$ flavors of vector-like fermion pairs $\chi_i$, $\bar{\chi}_i$, $i = 1\dots N_f$, assumed to transform as SM singlets and fundamentals of the $\lab{SU}(N)$.

CP is assumed to be an exact symmetry of the ultraviolet theory, with the invariant topological angle of the hidden sector fixed to $\bar{\theta}' = \pi$.
Below the confinement scale, CP is spontaneously broken by the chiral condensate, 
\begin{equation}
\langle \chi_i \bar{\chi}_i \rangle \sim \LCP^3 \e^{i\eta}
\end{equation}
where $\eta$ is a phase that acquires a vacuum expectation value. We assume that this condensate is the {\em only} source of CP violation in the theory. 
If the hidden sector consisted only of $\lab{SU}(N)$ gauge dynamics with a single fermion $\chi$, spontaneous CP breaking would require $\Lambda > m_\chi \gtrsim \Lambda/N$, implying a significantly large $N$. However a flavor symmetry of the multiplet relaxes this condition to 
\be
\frac{\delta m_\chi}{m_\chi} \lesssim \frac{N m_\chi}{\LCP}.
\label{eq:split}
\ee 
Therefore a good flavor symmetry alleviates the hierarchy between $m_\chi$ and $\LCP$. The flavor symmetry will be discussed shortly, and the condition (\ref{eq:split}) is derived in Appendix~\ref{app:cp}.

The details of the model lie in how CP violation is transmitted to the Standard Model in such a way that an $\mathcal{O}(1)$ CKM phase is obtained while keeping $\thetabar_{\textsc{qcd}} \approx 0$. 
For this, we consider the minimal field content elucidated in ref.~\cite{Bento:1991ez}. We augment the SM with a set of vector-like quarks $D$, $\bar{D}$ that transform like the right-handed down-type quarks of the SM. 
CP violation will be communicated to the SM via four-fermion operators $(\chi \bar{\chi}) (D \bar{d})$, generated via some heavy mediators at a scale $M \gg \LCP$.
In our minimal setup we take the mediator fields to consist of a single complex scalar, $\varphi$.

Our final assumption is that there exists a $\mathbb{Z}_2$ symmetry under which $\bar{\chi}$, $D$, $\bar{D}$ and $\varphi$ are odd while all other fields (including $\chi$ and all the SM matter) are even. 
The renormalizable, $\textrm{SM} \times \lab{SU}(N) \times \mathbb{Z}_2$-invariant interactions are then
\begin{equation}
\label{eq:uv_lagrangian}
\mathcal{L} ~\supset~ (m_{\chi})_{ab} \chi^a \bar{\chi}^b + \mu_D D \bar{D} - V(\varphi, \varphi^{\dagger})
+ (\lambda_{ab} \varphi + \tlambda_{ab} \varphi^{\dagger}) \chi^a \bar{\chi}^b 
+ (\kappa_i \varphi + \tkappa_i \varphi^{\dagger}) \bar{d}_i D
+ \hc 
\end{equation}
where $a, b = 1 \dots N_f$, and
where CP-invariance allows us to work in a basis where $\lambda, \tlambda, \kappa, \tkappa$ are all real. 
The $\mathbb{Z}_2$ symmetry forbids a renormalizable $Q H^c \bar{D}$ operator, which is crucial to the Nelson--Barr mechanism, as discussed below.
We have included an explicit mass term for $\chi$, $\bar{\chi}$ that softly breaks the $\mathbb{Z}_2$ symmetry, but as we will discuss shortly, its presence is not crucial for the setup.

After the fermions condense, $\varphi$ obtains a parametrically small vacuum expectation value,
\begin{equation}
\langle \varphi \rangle \simeq (\lambda + \tlambda) \frac{\LCP^3}{M^2}\, ,
\end{equation}
where $M$ is the mass of the scalar, $V(\varphi) \supset M^2 \varphi^{\dagger}\varphi$. 
This spontaneously breaks the $\mathbb{Z}_2$ symmetry and leads to an additional mass term for $\chi$, $\bar{\chi}$.
Since the spontaneous breaking of the $\mathbb{Z}_2$ does generate a fermion mass, the explicit $\mathbb{Z}_2$ breaking fermion mass is not needed at all. Hence we can take the $\mathbb{Z}_2$ to be an exact symmetry, assuming that explicit soft-breaking mass terms are parametrically smaller than the effective mass generated by the scalar vacuum expectation value, $m_{\chi} \approx \lambda \langle \varphi \rangle$. In this limit, the requirement for spontaneous CP breaking in Eq.~\eqref{eq:split} translates directly into a constraint on the flavor structure of the Yukawa couplings. Substituting the dynamically generated mass into Eq.~\eqref{eq:split}, we estimate that CP is spontaneously broken provided the flavor splitting in the couplings satisfies
\be
\label{eq:delta_lambda_req}
\frac{\delta \lambda }{\lambda} \lesssim N \lambda^2 \frac{\LCP^2}{M^2}.
\ee
To satisfy this condition, we assume that the Yukawa couplings preserve the $\lab{SU}(N_f)$ flavor symmetry of the hidden sector, such that $\lambda_{ab} \propto \delta_{ab}$. This assumption is analogous to our treatment of the exact $\mathbb{Z}_2$ symmetry; however, unlike the discrete $\mathbb{Z}_2$ parity, there are no irrelevant operators that can spontaneously break this continuous symmetry. Furthermore, the suppression of higher-dimensional operators that might explicitly break this flavor symmetry can be dynamically justified if the hidden gauge sector exhibits walking dynamics, similar to ``walking technicolor'' models~\cite{Holdom:1984sk, Yamawaki:1985zg, Appelquist:1986an}. In such scenarios, the large anomalous dimensions of the operators in the conformal window can sufficiently suppress flavor-violating effects from the UV, preserving the mass degeneracy required for spontaneous CP breaking.

It is worth emphasizing that, in contrast to ``ordinary'' Nelson--Barr models in which a scalar coupling to $D \bar{d}$ acquires a complex vacuum expectation value of $\mathcal{O}(M)$, we are assuming that $\varphi$ does {\em not} obtain a vacuum expectation value from the scalar potential, and does so only via its coupling to the $\lab{SU}(N)$ condensate. This is the essential point that ameliorates the fine-tuning issues discussed below. This also leads to a less severe electroweak hierarchy problem in the theory, as operators coupling $\varphi$ to the SM Higgs, $b (\varphi^{\dagger}\varphi) H^\dagger H$ contribute only a parametrically suppressed correction to the Higgs mass parameter below the scale of spontaneous CP breaking, $\sim \lambda (\LCP^6 / M^4) H^{\dagger}H$. 
The amount of tree-level fine-tuning in the electroweak sector is thus suppressed by $(\LCP / M)^2$ compared to the original model in \cite{Bento:1991ez}. 

Integrating out the complex scalar leads to a number of four-fermion operators, including
\begin{equation}
\label{eq:four_fermion_operators}
\mathcal{L} \supset \frac{1}{M^2}\big[ (\kappa_i \tlambda + \tkappa_i \lambda) (\chi \bar{\chi}) + (\kappa_i \lambda + \tkappa_i \tlambda) (\chi^{\dagger}\bar{\chi}^{\dagger}) \big] D \bar{d} + \hc 
\end{equation}
which, after the fermions condense, leads to an effective mass term for $D \bar{d}$ that carries the CP-violating phase. These effective interactions are essentially the same as those realized in ref.~\cite{Vecchi:2025qie} as well.

At low energies, the resulting mass matrix for the down-type quarks in the SM can be written
\begin{equation}
\label{eq:def_M0}
\mathcal{L} \supset \bp Q & D \ep \mathcal{M}_0 \bp \bar{d} \\ \bar{D} \ep\,, 
\quad \text{with} \quad 
\mathcal{M}_0 = \bp m_d & 0 \\ F & \mu_D \ep \,,
\end{equation}
where $(m_d)^i{}_j = (\lambda_d)^i{}_j v / \sqrt{2}$ and $F$ is a {\em row} vector with entries
\begin{equation}
\label{eq:f_values}
F_i = \frac{\LCP^3}{M^2} \big[ (\kappa_i \tlambda + \tkappa_i \lambda) \e^{i \eta}
+ (\kappa_i\lambda + \tkappa_i\tlambda) \e^{\minus i\eta} \big]\, .
\end{equation}
Here, the key point of the Nelson--Barr mechanism is manifest: as a result of the $\mathbb{Z}_2$ symmetry, the upper-right block of $\mathcal{M}_0$ vanishes. Thus, while $F$ has an $\mathcal{O}(1)$ CP-violating phase, the determinant of $\mathcal{M}_0$ is manifestly real so that $\bar{\theta} = 0$ at tree-level.
Note that it is crucial that the coefficients appearing in front of $\chi\bar{\chi}$ and the conjugate term in Eq.~\eqref{eq:four_fermion_operators} are not equal, or else $F$ would be real. 

We next need to understand under what conditions the mass matrix $\mathcal{M}_0$ leads to an $\mathcal{O}(1)$ CKM phase for the SM quarks. Here, we follow the discussion of ref.~\cite{Asadi:2022vys} closely (see also refs.~\cite{Bento:1991ez,Dine:2015jga,Perez:2020dbw}).
The mass matrix $\mathcal{M}_0$ can be diagonalized via the usual bi-unitary transformation $U_L^{\dagger} \mathcal{M}_0 U_R = \diag(\bar{m}, m_D)$, where $\bar{m}$ is the diagonal matrix of real, nonnegative SM quark masses and $m_D$ is the physical mass of the heavy, vector-like quarks. Assuming $m_D \gg \bar{m}$, the upper-left $3 \times 3$ block of $U_L$ can be identified as the usual CKM matrix, $V$ (we are working in the flavor basis in which the up-quark masses, omitted from Eq.~\eqref{eq:uv_lagrangian} for simplicity, are real diagonal).
Then, since $U_L^{\dagger} \mathcal{M}_0^{\dagger} \mathcal{M}_0 U_L = \diag(\bar{m}^2, m_D^2)$, we find
\begin{equation}
\label{eq:ckm_diagonalization}
(V \bar{m}^2 V^{\dagger})^{i}{}_j 
\simeq (m_d m_d^T)^i{}_j 
+ \frac{ (m_d)^i{}_k F^{\dagger\,k} F_l (m_d^T)^l{}_m }{F_p F^{\dagger\,p} + \mu_D^2}
\Bigg(\delta^m{}_j + \frac{(\bar{m}^2)^m{}_j}{F_q F^{\dagger\,q} + \mu_D^2} \Bigg)
\equiv 
(m_0^2)^i{}_j + \mathcal{O}(v^2 / \LCP^2)
\end{equation}
where in the last equality we have defined the effective $3 \times 3$ mass mixing matrix for the SM quarks,
\begin{equation}
\label{eq:effective_mixing_matrix}
(m_0^2)^i{}_j = (m_d)^i{}_k \bigg( \delta^k{}_l + \frac{F^{\dagger\,k} F_l}{F_p F^{\dagger\,p} + \mu_D^2}\bigg)(m_d^T)^l{}_j \, .
\end{equation}
Eq.~\eqref{eq:ckm_diagonalization} says that the effective mixing matrix for the SM quarks is approximately diagonalized by the CKM matrix. The upshot of this procedure is that, while the matrix $m_d$ is real, the effective mixing matrix is generally complex, due to the complex phase in $F_i$, and there are no ratios of $v / \LCP$ in Eq.~\eqref{eq:effective_mixing_matrix}. The phase in the CKM matrix will thus be $\mathcal{O}(1)$ if the two terms in parentheses in Eq.~\eqref{eq:effective_mixing_matrix} are of similar size. This in turn requires that $F F^{\dagger} \gtrsim \mu_D^2$, which translates to the condition
\footnote{On the other hand, ref.~\cite{Valenti:2021rdu} noted that if $FF^{\dagger} \gg \mu_D^2$, it is impossible to simultaneously avoid contributions to $\thetabar$ involving the vector-like quarks and to reproduce the SM Yukawas, due to their mixing. The resulting ``coincidence problem'' was an underlying motivation of the setup in ref.~\cite{Valenti:2021xjp}.}
\begin{equation}
\label{eq:ckm_condition}
\lambda^2 \kappa_i^2 \Big(\frac{\LCP}{M}\Big)^4 \gtrsim \Big( \frac{\mu_D}{\LCP} \Big)^2 \, .
\end{equation}
Here, we are assuming $\lambda \sim \tlambda$ and $\kappa_i \sim \tkappa_i$ for simplicity. 

There is an additional requirement to obtain an $\mathcal{O}(1)$ CKM phase, emphasized by ref.~\cite{Dine:2015jga}: the denominator $F_q F^{\dagger\,q} + \mu_D^2$ in Eq.~\eqref{eq:effective_mixing_matrix} is manifestly real, but we must ensure that the matrix in the numerator $F^{\dagger\,k} F_l$ has an $\mathcal{O}(1)$ phase. If only one of the $F_k$ is nonzero, or if each entry of $F_k$ has the same phase, the CKM matrix will be real. This is the reason that more than one real scalar degree of freedom acting as a portal is necessary. 

In summary, provided the above conditions including Eq.~\eqref{eq:ckm_condition} are satisfied, the strong CP problem is solved at tree-level. Thus far, other than the hierarchy problem associated with the new scale, the picture is quite similar to the minimal model in ref.~\cite{Bento:1991ez}. The utility of having the CP-violation transmitted through the chiral condensate will be made more clear when we consider higher-order corrections to this setup.

\subsection{Loop Corrections}
\label{subsec:loops}

While the structure of $\mathcal{M}_0$ in Eq.~\eqref{eq:def_M0} permits an $\mathcal{O}(1)$ CKM phase while keeping $\thetabar = 0$ at tree-level, this structure is spoiled by radiative corrections and higher-dimension operators.

To track these corrections, we follow ref.~\cite{Asadi:2022vys} and expand the down-quark mass matrix $\mathcal{M}_D = \mathcal{M}_0 + \mathcal{M}_1$, where $\mathcal{M}_1$ includes higher-order corrections, i.e., all entries of $\mathcal{M}_0^{-1} \mathcal{M}_1 \ll 1$. We then have
\begin{equation}
\label{eq:detMD_expand}
\det \mathcal{M}_D 
= \det \mathcal{M}_0 \times \det\big( \mathbbm{1} + \mathcal{M}_0^{-1} \mathcal{M}_1 \big) 
\simeq \det \mathcal{M}_0 \Big[ 1 + \tr\big( \mathcal{M}_0^{-1} \mathcal{M}_1 \big) \Big] 
\end{equation}
The corrections to $\thetabar$ are thus,
\begin{equation}
\Delta\thetabar \simeq \arg \Big[ 1 + \tr \big( \mathcal{M}_0^{-1} \mathcal{M}_1 \big) \Big] \simeq \Im \Big[ \tr \big( \mathcal{M}_0^{-1} \mathcal{M}_1\big) \Big] \, .
\end{equation}
We can write $\mathcal{M}_1$ in block form as
\begin{equation}
\label{eq:def_M1}
\mathcal{M}_1 = \begin{pmatrix}
\mathcal{M}_{Q\bar{d}}^{(1)} & \mathcal{M}_{Q\bar{D}}^{(1)} \\
\mathcal{M}_{D\bar{d}}^{(1)} & \mathcal{M}_{D\bar{D}}^{(1)} \end{pmatrix}\,, 
\end{equation}
so that
\begin{equation}
\label{eq:delta_thetabar}
\Delta\thetabar \simeq
\Im \Big( m_d^{-1} \mathcal{M}_{Q\bar{d}}^{(1)} - \frac{1}{\mu_D}
\big( F m_d^{-1} \mathcal{M}_{Q\bar{D}}^{(1)} + \mathcal{M}_{D\bar{D}}^{(1)} \big)\Big) \, .
\end{equation}
Note that corrections to the $\mathcal{M}_{D\bar{d}}^{(1)}$ mass term do not appear in this expression. In principle, these corrections can still spoil the Nelson--Barr mechanism if they dominate over the tree-level contribution to $F$ and suppress the phase in the CKM matrix. In the present model, however, these contributions are always suppressed (c.f., the variation of this setup with the CP violation in the glueball condensate in ref.~\cite{Perez:2020dbw}), so we will not consider them further.

The one-loop corrections depend on the form of the scalar potential for $\varphi$ and $H$. We first consider only the $\mathbb{Z}_2$-preserving terms,
\begin{equation}
V(\varphi, H) \supset  b_1 (\varphi^\dagger \varphi) H^{\dagger}H + c_1 (\varphi^{\dagger}\varphi)^2 + \Big( b_2 \varphi^2 H^{\dagger}H + c_2 \varphi^4 + c_3 \varphi^{\dagger} \varphi^3 + \hc \Big) 
\end{equation}
The only relevant assumptions on other terms in the potential is that they lead to a stable vacuum for $H$, but do not give a tree-level vacuum expectation value for $\varphi$.

\begin{figure}
\centering
\includegraphics[width=7cm]{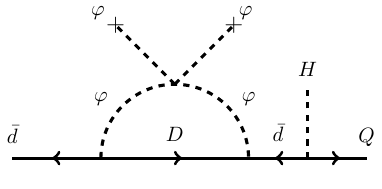}
\qquad 
\includegraphics[width=7cm]{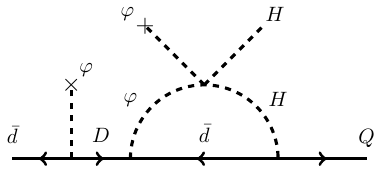} \\[0.5em]
\includegraphics[width=7cm]{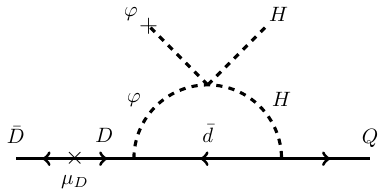}
\qquad
\includegraphics[width=7cm]{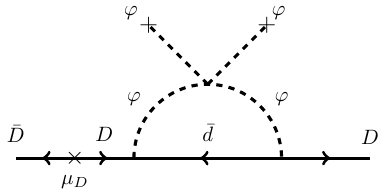}
\caption{One-loop diagrams showing the schematic corrections to $\mathcal{M}_{Q\bar{d}}$ (top row), $\mathcal{M}_{Q\bar{D}}$ (bottom-left) and $\mathcal{M}_{D\bar{D}}$ (bottom-right), leading to corrections to $\thetabar$.} 
\label{fig:diagrams_oneloop}
\end{figure}

Including these interactions, there are one-loop corrections to $\mathcal{M}_{Q\bar{d}}$, $\mathcal{M}_{Q\bar{D}}$ and $\mathcal{M}_{D\bar{D}}$, illustrated by the diagrams in Fig.~\ref{fig:diagrams_oneloop}. 
As an example, the upper-left diagram leads to the one-loop correction,\footnote{
Strictly speaking, we are computing finite, 1PI corrections to the dimension-five operator $\varphi^2 \bar{d}_j i \sigma^{\mu} D_{\mu} \bar{d}_i^{\dagger}$.
This is related to the operator $\varphi^2 Q_i H^c \bar{d}_j$ via the equations of motion for $\bar{d}$, and subsequently contributes to $\mathcal{M}_{Q\bar{d}}$ after $\varphi$ obtains a vacuum expectation value.}
\begin{equation}
\label{eq:oneloop_example}
\big(\mathcal{M}_{Q\bar{d}}\big)^i{}_j = \frac{1}{16\pi^2} c_3 \kappa^i \tkappa_k (m_d)^k{}_j \frac{\langle \varphi \rangle^2}{M^2} 
\end{equation}
The other diagrams are evaluated similarly, and---taking $\tlambda \sim \lambda$, $\tkappa \sim \kappa$, and all the quartic couplings $b_a, c_a \sim c$---the one-loop corrections all lead to the same order contribution to $\thetabar$:
\begin{equation}
\label{eq:delta_thetabar_oneloop}
\Delta\thetabar = \frac{1}{16\pi^2} c\,(\kappa \lambda)^2 \Big(\frac{\LCP}{M}\Big)^6
\end{equation}
This result could have been anticipated by noting that CP-violation arises only at the loop level via $\langle \varphi \rangle$, and that the $\mathbb{Z}_2$ symmetry requires two insertions of this expectation value.

It is clear from Eq.~\eqref{eq:delta_thetabar_oneloop} that even a modest hierarchy between $\LCP$ and $M$ will be enough to ensure that the solution to the strong CP problem is stable under quantum corrections.
This is in contrast to minimal models (c.f. \cite{Asadi:2022vys}) where a significant suppression of the quartic couplings (e.g., $b$ in our notation) is required to preserve $\thetabar \lesssim 10^{-10}$, which demands additional ingredients such as supersymmetry.  

Finally, we should note that in principle, operators that softly break the $\mathbb{Z}_2$ symmetry in the potential would lead to additional one-loop contributions as well, with slightly different scaling than Eq.~\eqref{eq:delta_thetabar_oneloop}. For instance, the operator $a \varphi H^{\dagger}H$ leads to a one-loop contribution to $\mathcal{M}_{Q\bar{D}}$ with no insertions of $\langle \varphi \rangle$.
However, treating the dimensionful coupling $a$ as a spurion of the $\mathbb{Z}_2$, it is natural to suppose that $a \sim m_{\chi} < \langle \varphi \rangle$ (the last inequality being an assumption we made in Section~\ref{subsec:setup}), in which case this contribution is subdominant to the one computed in Fig~\ref{fig:diagrams_oneloop}.

\subsection{Corrections from Higher-Dimensional Operators}

Aside from loop corrections, the Nelson--Barr solution to the strong CP problem can also be destabilized by non-renormalizable operators that spoil the texture in Eq.~\eqref{eq:def_M0}.
As discussed in the introduction, dimension-five operators already lead to a severe upper bound on $\LCP$ in the minimal BBP model, presenting a number of model-building challenges.

In the present case, when the only source of CP violation is in the chiral condensate, the quality problem is much less severe.
The only $\mathbb{Z}_2$-invariant operators that arise at dimension-five are
\begin{equation}
\label{eq:lag_dim5}
\mathcal{L} \supset 
\frac{h}{\Left} \varphi^{\dagger}\varphi D \bar{D} 
+ \frac{h'^i}{\Left} \varphi^{\dagger} Q_i H^c \bar{D}
+ \hc 
\end{equation}
where $h$, $h'$ are dimensionless, $\mathcal{O}(1)$ couplings and $i$, as usual, is a flavor index.

Treating these corrections as contributions to $\mathcal{M}^{(1)}$ in Eq.~\eqref{eq:def_M1}, the first operator leads to
\begin{equation}
\label{eq:nonrenorm_correction}
\Delta\thetabar \simeq \frac{h \langle \varphi^{\dagger}\varphi \rangle}{\mu_D \Left} \simeq \lambda^2\,h \frac{\LCP}{\Left} \frac{\LCP^4}{M^4} \frac{\LCP}{\mu_D}
\end{equation}
In the minimal BBP model, by contrast, the correction to $\thetabar$ is $\mathcal{O}(\LCP^2 / \mu_D \Lambda_{\textsc{eft}})$~\cite{Asadi:2022vys}.
In comparison, the correction in \eqref{eq:nonrenorm_correction} is suppressed by $\LCP^4 / M^4$. A modest hierarchy between the CP-breaking scale and the scalar mass $M$ thus allows for a much larger value for $\LCP$, even if $\mu_D$ is at much lower scales.

Accounting for the corrections from the second operator requires a more precise Ansatz for the flavor hierarchies in the couplings $h'^i$ and $\kappa_j$. We find,
\begin{equation}
\Delta\thetabar \simeq 
\lambda^2 \kappa_i (\lambda_d^{-1})^i{}_j h'^j \frac{\LCP}{\Lambda_{\textsc{eft}}} \frac{\LCP^4}{M^4} \frac{\LCP}{\mu_D}
\end{equation}
The combination $h'^i \kappa_j$ has the same flavor structure as the down-quark mass matrix $m_d$. Assuming these couplings have the same hierarchies as the down-type Yukawas (i.e., minimal flavor violation), the overall prefactor is $\mathcal{O}(1)$ and the correction has the same scaling as Eq.~\eqref{eq:nonrenorm_correction} (see the similar discussion in ref.~\cite{Asadi:2022vys}). 

Beyond dimension-five, there are further corrections from operators such as $(\chi \bar{\chi}) Q_i H^c \bar{D}$. These are suppressed by three powers of the cutoff, and thus always subdominant to the dimension-five operators above. 

The corrections in Eq.~\eqref{eq:nonrenorm_correction} present an immediate tension with the condition in Eq.~\eqref{eq:ckm_condition}: for a fixed hierarchy between $\LCP$ and $M$, a small vector-like quark mass $\mu_D \ll \LCP$ is preferred to achieve an $\mathcal{O}(1)$ phase in the CKM matrix, but this {\em enhances} the effects of non-renormalizable operators.
In the minimal model discussed here, the $(\LCP / M)^4$ suppression in \eqref{eq:nonrenorm_correction} is enough to ensure ample parameter space with a stable solution to the strong CP problem, even with $\mathcal{O}(1)$ values of $\lambda$ and $\kappa$, but the relationship between $\mu_D$, $\LCP$ and $M$ is relatively constrained. One could easily imagine additional symmetries that forbid operators to higher dimensions, {\em a la} ref.~\cite{Asadi:2022vys}, but we will not rely on these here.
There is also a tension between the need for a separation between $\LCP$ and $M$ and the resulting tuning required in the dark quark Yukawas, Eq.~\eqref{eq:delta_lambda_req}. 
For $\LCP \sim 10^8~\textrm{GeV}$, this tuning can be as small as $10^{-2}$, with $\LCP / M$ still small enough that the contributions from the dimension-5 operators are $< 10^{-10}$. This is roughly the value of $\LCP$ for which the quality problem is alleviated in a ``vanilla'' Nelson--Barr model~\cite{Dine:2015jga}, though in our case the loop corrections and electroweak hierarchy problem are comparatively less severe. Larger values of $\LCP$ can be easily accommodated in our model at the cost of tuning the Yukawas (or additional model-building for the necessary flavor symmetry).

\bigskip

In summary, we see that allowing for CP violation to arise only from a hidden sector condensate automatically resolves a number of the issues with the minimal Nelson--Barr models in a theoretically compelling manner. The amount of tuning required in the model is quite modest, and the parameter space with which the strong CP problem is successfully resolved is substantial, with significant freedom to have CP spontaneously broken up to $\LCP \lesssim 10^{13}\,\textrm{GeV}$, or down to scales where direct phenomenological constraints start to play a role. 
In the next section, we will show that an entirely orthogonal consideration can make a sharper prediction about the scales in the model, when the hidden sector also provides a candidate for the dark matter.

\section{Freeze-In Production from the CP Portal}
\label{sec:dm_abundance}

Beyond addressing the strong CP problem, this model can also account for the origin of dark matter without introducing any additional degrees of freedom. 
In this sense, it shares the same appealing features as the axion solution, which simultaneously solves the strong CP problem and provides a dark matter candidate. 

After chiral symmetry breaking at the confinement scale $\LCP$, the pseudo-Nambu--Goldstone bosons (PNGBs), referred to as dark pions $\tilde\pi$, emerge with masses given by $m_{\tilde\pi}^2 = m_\chi B_{\tilde\pi}$, where $B_{\tilde\pi} \sim 4\pi\LCP$. 
Since $m_\chi \gtrsim \LCP^3 / M^2$ and thus $m_{\tilde\pi} \gtrsim \LCP^2 / M$ (where $m_{\chi}$ here is the effective fermion mass including the contribution from the $\varphi$ vev), it is reasonable to consider $m_{\tilde\pi} \gg \mu_D$. 

If the chiral flavor symmetry is exact---i.e., in the absence of explicit mass splittings or portal interactions that distinguish the individual dark fermion flavors---the dark pions are absolutely stable.
To see this, note that the dark pions transform as an adjoint representation under the diagonal flavor symmetry $\mathrm{SU}(N_f)_V$ (under which $\chi$, $\bar{\chi}$ simultaneously transform as conjugates), so no flavor-singlet operator involving a single dark pion can be constructed.\footnote{Even when explicit breaking reduces the full flavor group, an unbroken residual symmetry (for example, a $\lab{U}(1)$ subgroup) can render the dark pion charged under it cosmologically stable.  However, one must still ensure that this charged pion is the lightest dark‑pion state; otherwise, charge‑conserving annihilations (e.g.\ $\tilde\pi^+ \tilde\pi^-\to\tilde\pi^0 \tilde\pi^0$) will deplete the charged relic into neutral pions, which promptly decay.  Moreover, even if the charged pion is the lightest, the cosmological evolution can be nontrivial due to up‑scattering into (decaying) heavier dark pions, whose rates are exponentially suppressed by their mass splittings. }
In what follows, we will assume that the chiral flavor symmetry is exact and investigate the dark pion as a viable dark matter candidate. While an exact symmetry is not required for spontaneous CP violation, it is clearly advantageous in light of Eq.~\eqref{eq:delta_lambda_req} and significantly opens the parameter space in which the strong CP problem is simultaneously solved.

Turning to the relic abundance, given the significant hierarchy between scales $M$ and $\LCP$, thermal freeze-out is ineffective. Consequently, if dark pions make up the dark matter, they must be produced through a different mechanism. In this work, we focus on freeze-in~\cite{Hall:2009bx}.

Assuming hereon that $\tkappa \sim \kappa$ and $\tlambda \sim \lambda$ to simplify the notation, the relevant interaction terms in the effective chiral Lagrangian are
\begin{equation}
\kappa_i \lambda\, 
\frac{F_{\tilde\pi}^2 B_{\tilde\pi}}{M^2}\Tr\left(e^{\sqrt{2} i \tilde\pi^j T^j/F_{\tilde\pi}}\right) D \bar{d}_i,
\label{eq:chptlag}
\end{equation}
where $F_{\tilde\pi}$ is the decay constant of the dark pions and $T_j$ are the generators of the $\lab{SU}(N_f)$ flavor symmetry. The dominant freeze-in production channel is via the annihilation of $D$ and $\bar{d}_i$ into two dark pions:
\begin{equation}
D + \bar{d} \to \tilde\pi + \tilde\pi.
\end{equation}
The Boltzmann equation describing this production is
\begin{equation}
\dot{n}_{\tilde\pi} + 3n_{\tilde\pi} H \approx 3 \int \dd\Pi_D \dd\Pi_{\bar{d}}\dd\Pi_1 \dd\Pi_2 (2\pi)^4 \delta^4 \left(p_D + p_{\bar{d}} - k_1 - k_2 \right) |M|^2_{D \bar{d} \to 2\tilde\pi} f_D f_{\bar{d}},
\end{equation}
where the factor of 3 accounts for quark flavors, and $f_i$ is the phase-space distribution of particle $i$. Assuming the masses of $D$ and $\bar{d}$ are negligible compared to $m_{\tilde\pi}$, and approximating $f_i$ as Maxwell--Boltzmann distributions, the Boltzmann equation simplifies to~\cite{Edsjo:1997bg}
\begin{equation}
\dot{n}_{\tilde\pi} + 3n_{\tilde\pi} H \approx \frac{3T}{514\pi^5} \int^{\infty}_{4m_{\tilde\pi}^2} \dd s\, \dd\Omega \sqrt{s - 4m_{\tilde\pi}^2}\, |M|^2_{D \bar{d} \to 2\tilde\pi} K_1(\sqrt s / T),
\end{equation}
where $K_1$ is the modified Bessel function of the second kind. The squared matrix element is 
\begin{equation}
\sum |\mathcal{M}|^2_{D\bar{d}\to 2\tilde\pi} = 
12 |\kappa \lambda |^2 
\frac{B_{\tilde\pi}^2}{M^4} s \, . 
\end{equation}

Note that the production rate of this freeze-in process scales approximately as $T^6/M^2$, indicating that it is UV-dominated and therefore sensitive to the reheating temperature $T_\mathrm{rh}$. However, the reheating temperature cannot be arbitrarily high: it must remain below $\LCP$ to prevent the formation of a CP domain wall~\cite{McNamara:2022lrw, Asadi:2022vys}.
Integrating the Boltzmann equation we obtain the approximate dark pion yield (using the definitions of the yield $Y = n/s$ and the expression of the entropy density $s = 2\pi^2 g_*^s T^3/45$): 
\begin{equation}
Y_{\tilde\pi} \approx 0.02 \frac{N_{\tilde{\pi}}}{g_*^s \sqrt{g_*^\rho}} \frac{M_{\textrm{Pl}} B_{\tilde\pi}^2}{M^4} 
| \kappa \lambda |^2\, T_\mathrm{rh}
\end{equation}
with $N_{\tilde\pi} = N_f^2 - 1$ representing the number of (stable) dark pion species. Here $g_*^{s,\rho}$ are the effective numbers of degrees of freedom in the bath at the freeze-in temperature for the entropy $S$ and energy density $\rho$ respectively.
To obtain the correct relic density $\Omega_{\tilde{\pi}} h^2 \simeq 0.12$, we set
\begin{equation}
\Omega_{\tilde\pi} h^2 \approx \frac{2 m_{\tilde\pi} Y_{\tilde\pi}}{3.6\times 10^{-9} \gev} \simeq 0.12.
\end{equation}
Under the assumption that $m_\chi$ is entirely generated by $\langle\varphi\rangle$, we derive the scale hierarchy between $\LCP$ and $M$ required to achieve the correct relic density via freeze-in:
\begin{equation}
\label{eq:relic_density_scale}
\frac{\LCP}{M} \simeq 10^{-5} \left(\frac{4\pi}{B_{\tilde\pi}/\LCP}\right)^{1/2}
\left(\frac{10}{T_\mathrm{rh}/m_{\tilde\pi}}\right)^{1/6}
\left(\frac{3}{N_{\tilde\pi}}\right)^{1/4}\left(\frac{1}{\kappa^{2/3}\lambda}\right)
.
\end{equation}
Clearly, smaller values of the parameters $B_{\tilde\pi}$, $T_\mathrm{rh}$ and $\kappa, \lambda$ lead to larger values of $\LCP$, as the freeze-in mechanism requires a suppressed production rate.
On the other hand, a larger number of channels $N_{\bar{\pi}}$ enhances the production, thus lowering the required $\LCP$.
Note that, because this dark matter candidate is very heavy, self-interaction constraints are negligible.

\medskip

\begin{figure}[htp!]
\centering
\includegraphics[width=0.8\linewidth]{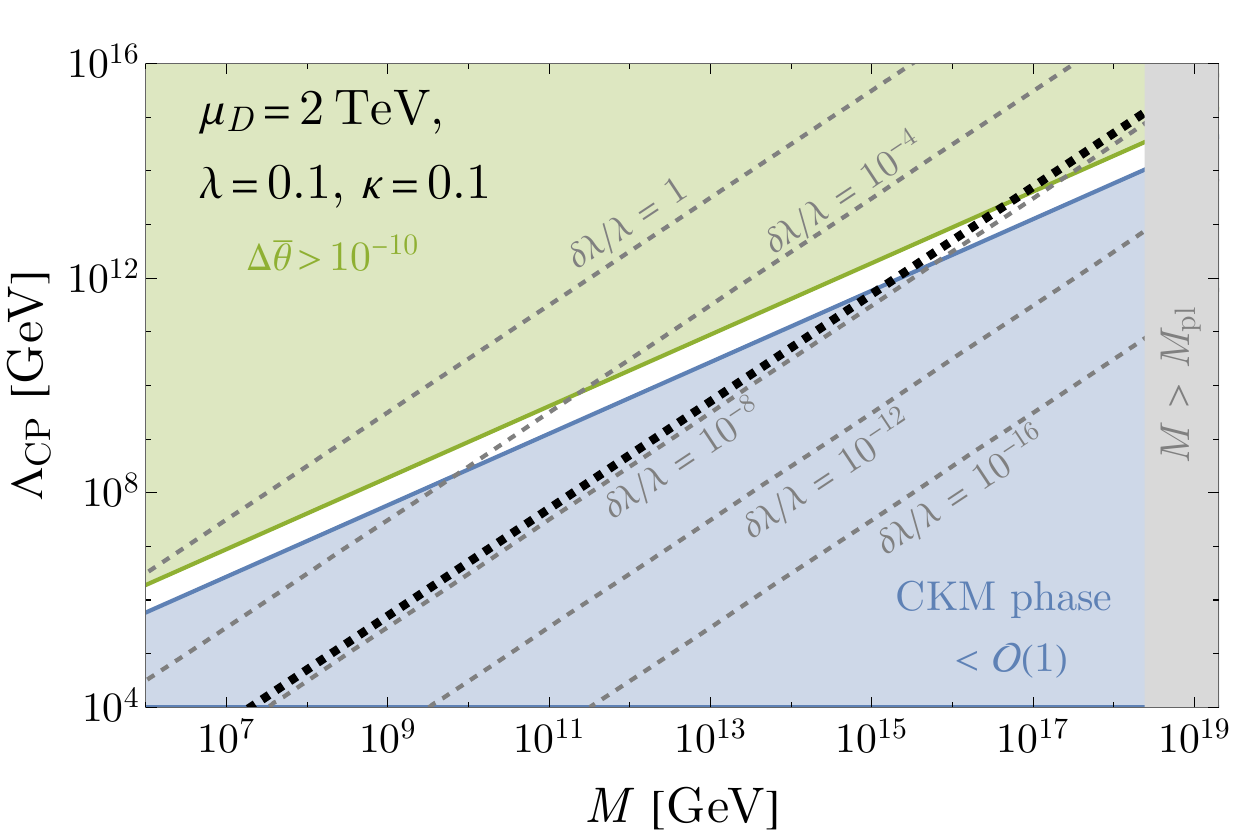}\\
\includegraphics[width=0.8\linewidth]{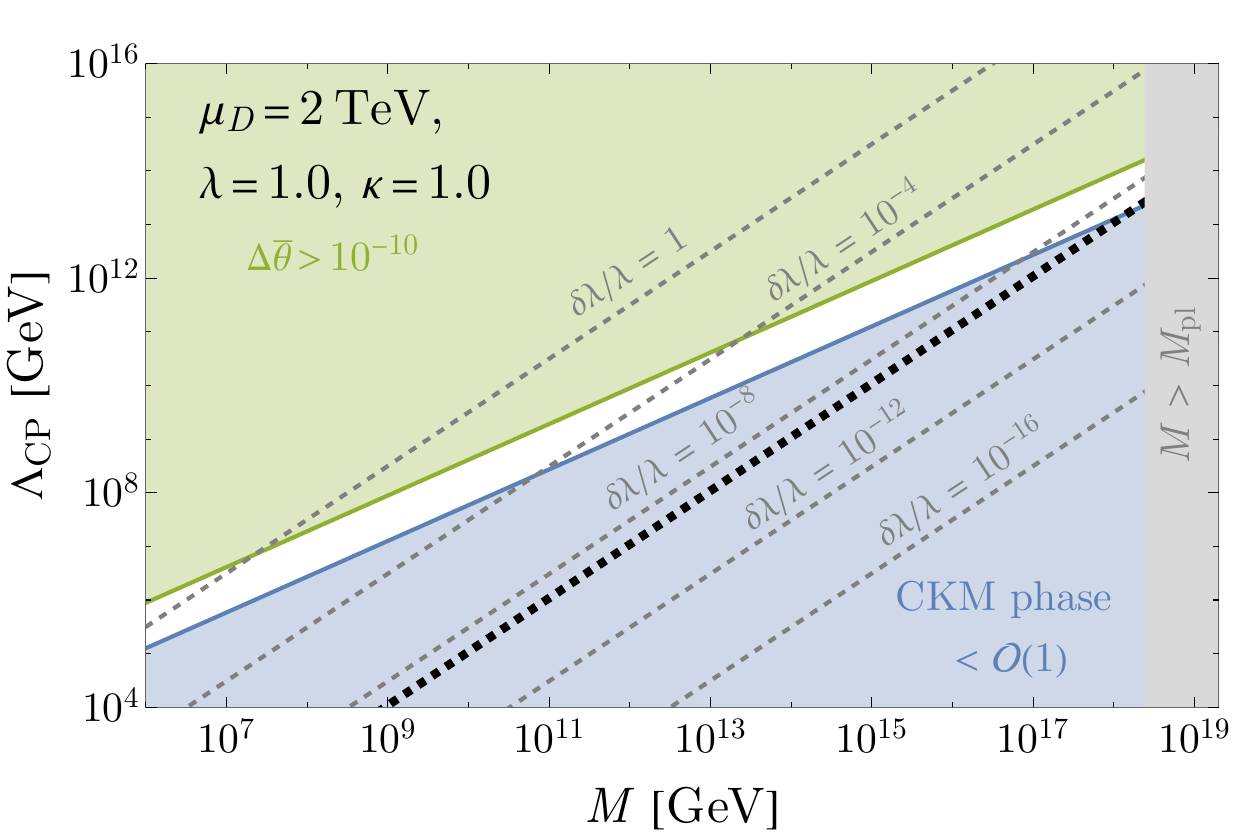}\\
\vskip -0.25em
\caption{
Region plots showing the scales in the model for which the strong CP and dark matter problems are successfully solved via coupling to a strongly-coupled $\lab{SU}(N)$ sector with $N_f = 2$ dark quarks. 
In these panels we fix $\mu_D = 2\,\textrm{TeV}$ and show the constraints as a function of $\LCP$ and $M$, with the correct freeze-out abundance indicated by the dashed black line. We also show contours of constant tuning for the dark quark Yukawa couplings, $\delta \lambda / \lambda$ assuming $\lab{SU}(10)$.
In the blue region, the CKM phase is too small to reproduce the SM while in the green region the corrections to $\thetabar$ from non-renormalizable operators are too large. The grey region indicates where $M$ is larger than the Planck scale. We assume $T_{\textrm{rh}} = 10\, m_{\tilde{\pi}}$.
}
\label{fig:region_plots_1}
\end{figure}

\begin{figure}[ht!]
\centering
\includegraphics[width=0.48\linewidth]{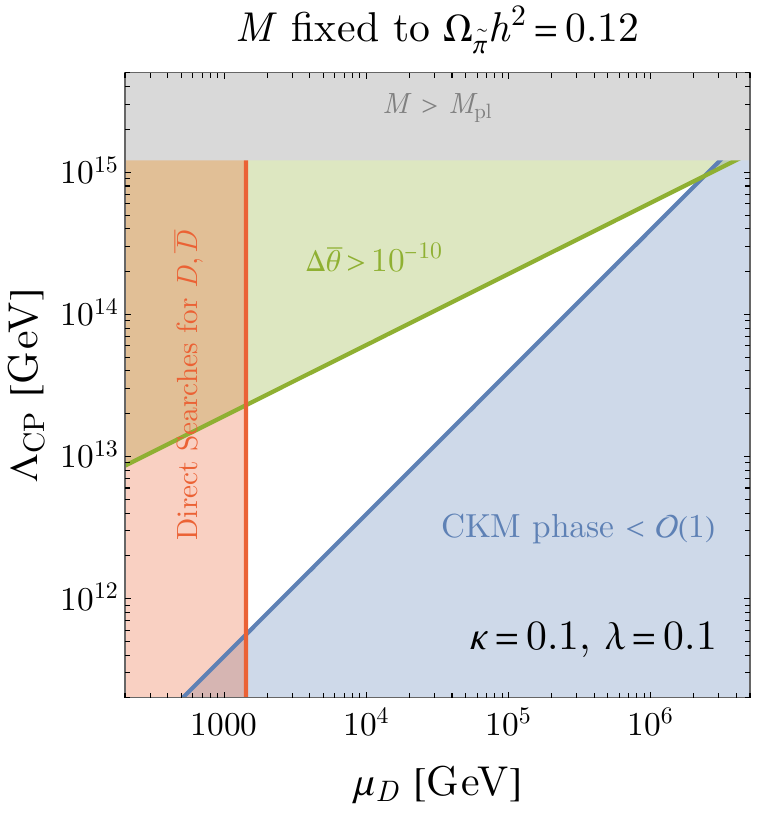}
~
\includegraphics[width=0.48\linewidth]{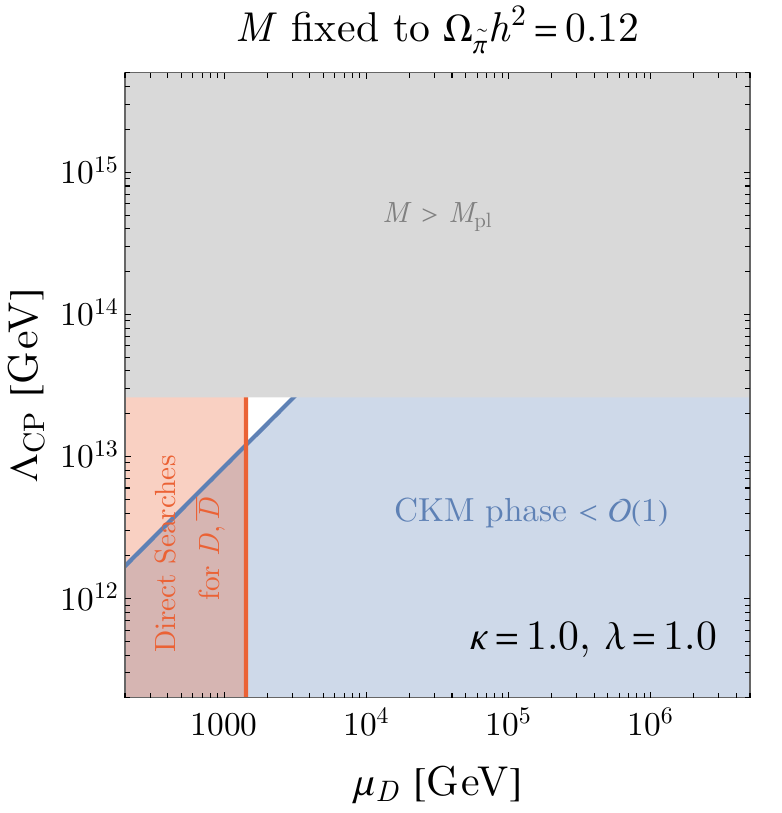}
\caption{Similar to Fig.~\ref{fig:region_plots_1}, except that we show the constraints as a function of $\LCP$ and $\mu_D$, with $M$ fixed to reproduce the correct relic abundance via freeze-out. 
The red region shows the direct LHC constraints on vector-like quarks, while the other regions are the same as in Fig.~\ref{fig:region_plots_1}.}
\label{fig:region_plots_2}
\end{figure}

With all the ingredients in place, we can search for regions of parameter space where the correct relic abundance is obtained while simultaneously satisfying the theoretical constraints outlined in Section~\ref{sec:model}. 
It's clear from Eq.~\eqref{eq:relic_density_scale} that, in order to obtain the correct relic density, we need several orders of magnitude splitting between $\LCP$ and $M$. From Eq.~\eqref{eq:ckm_condition}, this in turn requires that $\mu_D$ be at a much lower scale. 

With $\mu_D \ll \LCP$, the tension between Eq.~\eqref{eq:nonrenorm_correction} and \eqref{eq:ckm_condition} thus implies a strong constraint on the parameter space in which the Nelson--Barr mechanism is viable. Nevertheless, there is appreciable parameter space where the strong CP problem is solved and the freeze-in relic abundance is obtained.

This is demonstrated in Figs.~\ref{fig:region_plots_1} and \ref{fig:region_plots_2}. In Fig.~\ref{fig:region_plots_1}, we fix $\mu_D = 2~\textrm{TeV}$ and consider the various constraints on the model as a function of $\LCP$ and $M$. The region where the CKM phase is too small is shown in blue, while the region where the corrections from dimension-five operators are too large is shown in green. The grey region is excluded by requiring $M$ to be sub-Planckian. The white region shows the parameter space which solves the strong CP problem. Here we also show contours of $\delta \lambda / \lambda$ from Eq.~\eqref{eq:delta_lambda_req}, assuming $N = 10$. The black dashed contour shows the parameters that give the correct relic density. We see that spontaneous CP violation requires a significant tuning of the Yukawas in all the parameter space where the relic density is obtained, independently of the requirement of a flavor symmetry for the dark pions to be stable.

In Fig.~\ref{fig:region_plots_2} we instead show the parameter space in the $\LCP$ vs. $\mu_D$ plane, with $M$ fixed to obtain the correct relic density. Here the red region is excluded by direct searches for the vector-like states at the LHC~\cite{ATLAS:2022hnn, ATLAS:2022tla}. The white region simultaneously solves the strong CP and dark matter problems. In the left panel the tuning $\delta \lambda / \lambda$ required by Eq.~\eqref{eq:delta_lambda_req} is everywhere $\sim 10^{-8}$ while in the right panel it is everywhere $\sim 10^{-9}$, again assuming $N = 10$.

One notable feature of the allowed parameter space is that smaller values of $\mu_D$---from the current constraints up to the few tens of TeV---tend to be preferred. This is in large part due to the condition Eq.~\eqref{eq:ckm_condition}: the correct relic abundance and suppression of fine-tuning issues require a moderate hierarchy between the scales $\LCP$ and $M$, and this is more easily obtained when $\mu_D \ll \LCP$. As a result, a large amount of the available parameter space is testable at future hadron colliders beyond the LHC.

\section{Discussion and Conclusions}
\label{sec:discussion}

In summary, we have implemented a minimal Nelson--Barr solution to the strong CP problem, where the only source of CP-violation is the spontaneous breaking by the confining dynamics of a strongly-coupled hidden sector.
We demonstrated that this origin of CP-violation naturally reduces the need for fine-tuning in the model and provides a degree of protection from non-renormalizable operators that would otherwise lead to unacceptable contributions to $\thetabar$. 

We also showed that this model provides a natural dark matter candidate in the dark pions of the hidden sector, whose abundance is fixed by the freeze-in mechanism using the same portal that is responsible for transmitting the CP-violation to the Standard Model quark sector. There is ample parameter space where the strong CP problem and dark matter abundances are simultaneously resolved, while satisfying all the existing constraints. Furthermore, this parameter space favors smaller masses for the vector-like matter extending the visible sector. This provides a compelling target for future, high-energy proton colliders.

While we have focused on a minimal implementation of this model, the features associated with having the CP violation originate from a chiral condensate could easily be implemented in more complicated theories. For instance, the $\lab{SU}(N)$ could straightforwardly be a part of a larger gauge group. 
One could also further suppress the contributions from higher-dimension operators by invoking additional symmetries (e.g., the gauged combination of baryon number and hypercharge in ref.~\cite{Asadi:2022vys}). 
This could significantly open the available parameter space, particularly in regions where the necessary tuning of the Yukawas from Eq.~\eqref{eq:delta_lambda_req} is less severe. 
The scalar portal could also be replaced with other dynamics to generate the four-fermion operators in Eq.~\eqref{eq:four_fermion_operators}; it would be interesting to study alternatives in which the hierarchy between $\LCP$ and $M$ is generated automatically.
It would also be interesting to study realizations where the assumed flavor symmetry in the dark $\lab{SU}(N)$ sector is generated dynamically, as mentioned in Section~\ref{subsec:setup}.

One notable barrier to further model-building is the challenge in constructing a supersymmetric version of the minimal model presented here. In a supersymmetric version of the present model, the $\tilde{\lambda}$ and $\tilde{\kappa}$ terms in Eq.~\eqref{eq:uv_lagrangian} are forbidden by holomorphy of the superpotential. As a result, there is only one term in Eq.~\eqref{eq:f_values}, and all the entries of $F_i$ have the same sign so there is no phase in the CKM matrix. Unlike minimal BBP models, this result is unchanged even if additional scalar degrees of freedom (with a $Z_N$ rather than a $Z_2$ symmetry) are added to the model; additional scalar mediators all communicate the same phase from the confining sector, and the CKM matrix remains real. A simple (if unsatisfying) resolution to this problem would be to add a second confining sector, but it would be interesting to explore other possibilities.

Another aspect of these theories that we have not explored is whether they can contain a mechanism for baryogenesis. This is particularly important because baryogenesis models require a {\em new} source of CP violation beyond the SM, and we have supposed that the only source is in the chiral condensate of the hidden sector. 
One could easily construct additional portals coupling to the phase of the chiral condensate, but any further coupling to the SM (which is presumably necessary to populate the SM abundances) would result in new, higher-order corrections to $\thetabar$ which must be checked.
As our model easily incorporates a mechanism for producing the dark matter, it would be even more interesting to see if an extension that includes baryogenesis would also explain the apparent $\mathcal{O}(1)$ coincidence between the cosmological abundances of baryonic and dark matter, as in refs.~\cite{Fukugita:1986hr, Farina:2016ndq, Feng:2020urb, Kilic:2021zqu}.

\section*{Acknowledgments}

We are grateful to Kevin Langhoff, Seth Koren, Matthew Reece and Scott Watson for useful discussions. We also thank the anonymous JHEP referee for important comments on the flavor symmetry. 
The authors are supported by the NSF grant PHY-2309456. CC and SH are also supported in part by the US-Israeli BSF grant 2016153. TY is also supported by the Samsung Science and Technology Foundation under Project Number SSTF-BA2201-06.

\appendix
\section{CP Phase at \texorpdfstring{$\bar\theta=\pi$}{thetabar = pi}}
\label{app:cp}

In this appendix, we review spontaneous CP breaking at $\thetabar = \pi$ using large-$N$ chiral perturbation theory ($\chi$PT). For our purposes, it suffices to focus on the cases with $N_f = 1$ and $N_f = 2$.

Assuming confinement and spontaneous chiral symmetry breaking induced by a quark–antiquark condensate at a generic value of $\bar\theta$ and in the large-$N$ limit, the effective potential for the single-flavor case takes the simple form~\cite{Rosenzweig:1979ay,DiVecchia:1980yfw,Nath:1979ik,Witten:1980sp}
\be
V_\chi \supset -m_q  \Sigma^3 \cos\phi + \frac{\chi_\mathrm{YM}}{2}(\bar\theta - \phi)^2,
\ee
where $\Sigma^3 = \langle \bar q q\rangle$ is the quark chiral condensate, $\phi$ is its phase, and $\chi_\mathrm{YM}$ is the topological susceptibility of pure YM theory in the large-$N$ limit.
Note that the quark mass $m_q$ is real as its phase is absorbed into $\bar\theta$.
The CP-breaking threshold at $\bar\theta = \pi$ occurs when the quadratic term about $\phi = \pi$ changes sign, yielding~\cite{DiVecchia:2017xpu}
\be
m_{Q,0} = \frac{\chi_\mathrm{YM}}{\Sigma^3} \sim \frac{\Lambda}{N}.
\ee
Now we consider the two-flavor case ($N_f=2$) with quark masses $m_u$ and $m_d$. We assume the ordering $m_u < m_d$. The effective potential for the chiral phases $\phi_u$ and $\phi_d$ is given by
\be
V(\phi_u, \phi_d) = -m_u \Sigma^3 \cos \phi_u - m_d \Sigma^3 \cos \phi_d + \frac{\chi_\mathrm{YM}}{2}(\bar\theta - \phi_u - \phi_d)^2.
\ee
The stationarity conditions $\partial V / \partial \phi_i = 0$ at $\bar\theta=\pi$ yield
\be
m_u \Sigma^3 \sin \phi_u = m_d \Sigma^3 \sin \phi_d = \chi_\mathrm{YM}(\pi - \phi_u - \phi_d).
\ee
A trivial CP-conserving solution exists at $\phi_u = \pi$ and $\phi_d = 0$. To determine the stability of this vacuum, we examine the mass matrix of the pions. A second-order phase transition to a CP-broken phase occurs when the lightest eigenstate becomes massless. This corresponds to the condition:
\be
\frac{1}{m_u} - \frac{1}{m_d} = \frac{ \Sigma^3}{\chi_\mathrm{YM}}.
\ee
When the difference on the left-hand side is smaller than the inverse topological susceptibility, the CP-conserving vacuum becomes unstable. Thus, the condition for spontaneous CP breaking is
\be
\frac{1}{m_u} - \frac{1}{m_d} < \frac{ \Sigma^3}{\chi_\mathrm{YM}}.
\ee
Defining the average mass $m_q \approx m_u \approx m_d$ and the mass splitting $\delta m_q = m_d - m_u$, and utilizing the large-$N$ scaling where $\chi_\mathrm{YM}/\Sigma^3 \sim \Lambda/N$, the condition for the CP-broken phase becomes
\be
\frac{\delta m_q}{m_q} \lesssim \frac{N m_q}{\Lambda}.
\label{eq:CPcondition}
\ee
This indicates that for sufficiently degenerate quark masses (or large $N$), the vacuum at $\bar\theta=\pi$ spontaneously breaks CP.
The effect of a heavy scalar field $\varphi$ coupled to the quarks is suppressed by its mass squared. As a result, the condition (\ref{eq:CPcondition}) for the spontaneous CP breaking is shifted at most by an $\mathcal O(1)$ factor.

{\small
\bibliography{chicond-nb-refs}

@article{Nelson:1983zb,
	author = {Nelson, Ann E.},
	doi = {10.1016/0370-2693(84)92025-2},
	journal = {Phys. Lett. B},
	pages = {387--391},
	reportnumber = {HUTP-83/A080},
	title = {{Naturally Weak CP Violation}},
	volume = {136},
	year = {1984}
	}

@article{Nelson:1984hg,
	author = {Nelson, Ann E.},
	doi = {10.1016/0370-2693(84)90827-X},
	journal = {Phys. Lett. B},
	pages = {165--170},
	reportnumber = {HUTP-84/A022},
	title = {{Calculation of $\theta$ Barr}},
	volume = {143},
	year = {1984}
	}

@article{Barr:1984qx,
	author = {Barr, Stephen M.},
	doi = {10.1103/PhysRevLett.53.329},
	journal = {Phys. Rev. Lett.},
	pages = {329},
	reportnumber = {DOE-ER-40048-08 P4},
	title = {{Solving the Strong CP Problem Without the Peccei-Quinn Symmetry}},
	volume = {53},
	year = {1984}
	}

@article{Barr:1984fh,
	author = {Barr, Stephen M.},
	doi = {10.1103/PhysRevD.30.1805},
	journal = {Phys. Rev. D},
	pages = {1805},
	reportnumber = {DOE/ER/40048-15 P4},
	title = {{A Natural Class of Non-Peccei-Quinn Models}},
	volume = {30},
	year = {1984}
	}

@article{Dine:2015jga,
	archiveprefix = {arXiv},
	author = {Dine, Michael and Draper, Patrick},
	doi = {10.1007/JHEP08(2015)132},
	eprint = {1506.05433},
	journal = {JHEP},
	pages = {132},
	primaryclass = {hep-ph},
	title = {{Challenges for the Nelson-Barr Mechanism}},
	volume = {08},
	year = {2015}
	}

@article{Bento:1991ez,
	author = {Bento, Luis and Branco, Gustavo C. and Parada, Paulo A.},
	doi = {10.1016/0370-2693(91)90530-4},
	journal = {Phys. Lett. B},
	pages = {95--99},
	reportnumber = {IFM-8-91},
	title = {{A Minimal model with natural suppression of strong CP violation}},
	volume = {267},
	year = {1991}
	}

@article{Harnik:2004su,
    author = "Harnik, Roni and Perez, Gilad and Schwartz, Matthew D. and Shirman, Yuri",
    title = "{Strong CP, flavor, and twisted split fermions}",
    eprint = "hep-ph/0411132",
    archivePrefix = "arXiv",
    reportNumber = "LBNL-56615, UCB-PTH-04-30",
    doi = "10.1088/1126-6708/2005/03/068",
    journal = "JHEP",
    volume = "03",
    pages = "068",
    year = "2005"
}

@article{McNamara:2022lrw,
    author = "McNamara, Jacob and Reece, Matthew",
    title = "{Reflections on Parity Breaking}",
    eprint = "2212.00039",
    archivePrefix = "arXiv",
    primaryClass = "hep-th",
    month = "11",
    year = "2022"
}

@article{Hiller:2001qg,
	archiveprefix = {arXiv},
	author = {Hiller, Gudrun and Schmaltz, Martin},
	doi = {10.1016/S0370-2693(01)00814-0},
	eprint = {hep-ph/0105254},
	journal = {Phys. Lett. B},
	pages = {263--268},
	reportnumber = {SLAC-PUB-8801, FERMILAB-PUB-01-037-T},
	title = {{Solving the Strong CP Problem with Supersymmetry}},
	volume = {514},
	year = {2001}
	}

@article{Hiller:2002um,
	archiveprefix = {arXiv},
	author = {Hiller, Gudrun and Schmaltz, Martin},
	doi = {10.1103/PhysRevD.65.096009},
	eprint = {hep-ph/0201251},
	journal = {Phys. Rev. D},
	pages = {096009},
	reportnumber = {SLAC-PUB-9080, FERMILAB-PUB-01-379-T, BUHEP-01-35},
	title = {{Strong Weak CP Hierarchy from Nonrenormalization Theorems}},
	volume = {65},
	year = {2002}
	}

@article{Perez:2020dbw,
    author = "Perez, Gilad and Shalit, Aviv",
    title = "{High quality Nelson-Barr solution to the strong CP problem with $\theta=\pi$}",
    eprint = "2010.02891",
    archivePrefix = "arXiv",
    primaryClass = "hep-ph",
    doi = "10.1007/JHEP02(2021)118",
    journal = "JHEP",
    volume = "02",
    pages = "118",
    year = "2021"
}

@article{Asadi:2022vys,
    author = "Asadi, Pouya and Homiller, Samuel and Lu, Qianshu and Reece, Matthew",
    title = "{Chiral Nelson-Barr models: Quality and cosmology}",
    eprint = "2212.03882",
    archivePrefix = "arXiv",
    primaryClass = "hep-ph",
    doi = "10.1103/PhysRevD.107.115012",
    journal = "Phys. Rev. D",
    volume = "107",
    number = "11",
    pages = "115012",
    year = "2023"
}

@article{Hall:2009bx,
    author = "Hall, Lawrence J. and Jedamzik, Karsten and March-Russell, John and West, Stephen M.",
    title = "{Freeze-In Production of FIMP Dark Matter}",
    eprint = "0911.1120",
    archivePrefix = "arXiv",
    primaryClass = "hep-ph",
    reportNumber = "OUTP-09-18-P, UCB-PTH-09-32",
    doi = "10.1007/JHEP03(2010)080",
    journal = "JHEP",
    volume = "03",
    pages = "080",
    year = "2010"
}

@article{Edsjo:1997bg,
    author = "Edsjo, Joakim and Gondolo, Paolo",
    title = "{Neutralino relic density including coannihilations}",
    eprint = "hep-ph/9704361",
    archivePrefix = "arXiv",
    reportNumber = "UUITP-11-97, MPI-PHT-97-27",
    doi = "10.1103/PhysRevD.56.1879",
    journal = "Phys. Rev. D",
    volume = "56",
    pages = "1879--1894",
    year = "1997"
}

@article{Dashen:1970et,
    author = "Dashen, Roger F.",
    title = "{Some features of chiral symmetry breaking}",
    doi = "10.1103/PhysRevD.3.1879",
    journal = "Phys. Rev. D",
    volume = "3",
    pages = "1879--1889",
    year = "1971"
}

@article{Gaiotto:2017tne,
    author = "Gaiotto, Davide and Komargodski, Zohar and Seiberg, Nathan",
    title = "{Time-reversal breaking in QCD$_{4}$, walls, and dualities in 2 + 1 dimensions}",
    eprint = "1708.06806",
    archivePrefix = "arXiv",
    primaryClass = "hep-th",
    doi = "10.1007/JHEP01(2018)110",
    journal = "JHEP",
    volume = "01",
    pages = "110",
    year = "2018"
}

@article{Gaiotto:2017yup,
    author = "Gaiotto, Davide and Kapustin, Anton and Komargodski, Zohar and Seiberg, Nathan",
    title = "{Theta, Time Reversal, and Temperature}",
    eprint = "1703.00501",
    archivePrefix = "arXiv",
    primaryClass = "hep-th",
    doi = "10.1007/JHEP05(2017)091",
    journal = "JHEP",
    volume = "05",
    pages = "091",
    year = "2017"
}

@article{DiVecchia:2017xpu,
    author = "Di Vecchia, Paolo and Rossi, Giancarlo and Veneziano, Gabriele and Yankielowicz, Shimon",
    title = "{Spontaneous $CP$ breaking in QCD and the axion potential: an effective Lagrangian approach}",
    eprint = "1709.00731",
    archivePrefix = "arXiv",
    primaryClass = "hep-th",
    reportNumber = "CERN-TH-2017-186, NORDITA-2017-093, TAUP-3023-17",
    doi = "10.1007/JHEP12(2017)104",
    journal = "JHEP",
    volume = "12",
    pages = "104",
    year = "2017"
}

@article{Cordova:2019jnf,
    author = "C\'ordova, Clay and Freed, Daniel S. and Lam, Ho Tat and Seiberg, Nathan",
    title = "{Anomalies in the Space of Coupling Constants and Their Dynamical Applications I}",
    eprint = "1905.09315",
    archivePrefix = "arXiv",
    primaryClass = "hep-th",
    doi = "10.21468/SciPostPhys.8.1.001",
    journal = "SciPost Phys.",
    volume = "8",
    number = "1",
    pages = "001",
    year = "2020"
}

@article{Cordova:2019uob,
    author = "C\'ordova, Clay and Freed, Daniel S. and Lam, Ho Tat and Seiberg, Nathan",
    title = "{Anomalies in the Space of Coupling Constants and Their Dynamical Applications II}",
    eprint = "1905.13361",
    archivePrefix = "arXiv",
    primaryClass = "hep-th",
    doi = "10.21468/SciPostPhys.8.1.002",
    journal = "SciPost Phys.",
    volume = "8",
    number = "1",
    pages = "002",
    year = "2020"
}

@article{Cordova:2019bsd,
    author = "C\'ordova, Clay and Ohmori, Kantaro",
    title = "{Anomaly Obstructions to Symmetry Preserving Gapped Phases}",
    eprint = "1910.04962",
    archivePrefix = "arXiv",
    primaryClass = "hep-th",
    month = "10",
    year = "2019"
}

@article{Smilga:1998dh,
    author = "Smilga, Andrei V.",
    title = "{QCD at theta similar to pi}",
    eprint = "hep-ph/9805214",
    archivePrefix = "arXiv",
    reportNumber = "TPI-MINN-98-08, ITEP-TH-25-98, TPI-MINN--98-08",
    doi = "10.1103/PhysRevD.59.114021",
    journal = "Phys. Rev. D",
    volume = "59",
    pages = "114021",
    year = "1999"
}

@article{Creutz:2003xu,
    author = "Creutz, Michael",
    title = "{Spontaneous violation of CP symmetry in the strong interactions}",
    eprint = "hep-lat/0312018",
    archivePrefix = "arXiv",
    doi = "10.1103/PhysRevLett.92.201601",
    journal = "Phys. Rev. Lett.",
    volume = "92",
    pages = "201601",
    year = "2004"
}

@article{Creutz:2006ts,
    author = "Creutz, Michael",
    title = "{One flavor QCD}",
    eprint = "hep-th/0609187",
    archivePrefix = "arXiv",
    doi = "10.1016/j.aop.2007.01.002",
    journal = "Annals Phys.",
    volume = "322",
    pages = "1518--1540",
    year = "2007"
}

@article{Draper:2018mpj,
    author = "Draper, Patrick",
    title = "{Domain Walls and the $CP$ Anomaly in Softly Broken Supersymmetric QCD}",
    eprint = "1801.05477",
    archivePrefix = "arXiv",
    primaryClass = "hep-th",
    reportNumber = "ACFI-T18-02",
    doi = "10.1103/PhysRevD.97.085003",
    journal = "Phys. Rev. D",
    volume = "97",
    number = "8",
    pages = "085003",
    year = "2018"
}

@article{Csaki:2024lvk,
    author = "Cs\'aki, Csaba and Ruhdorfer, Maximilian and Youn, Taewook",
    title = "{Spontaneous CP breaking in a QCD-like theory}",
    eprint = "2407.06252",
    archivePrefix = "arXiv",
    primaryClass = "hep-ph",
    doi = "10.1007/JHEP12(2024)066",
    journal = "JHEP",
    volume = "12",
    pages = "066",
    year = "2024"
}

@article{Murayama:2021xfj,
    author = "Murayama, Hitoshi",
    title = "{Some Exact Results in QCD-like Theories}",
    eprint = "2104.01179",
    archivePrefix = "arXiv",
    primaryClass = "hep-th",
    doi = "10.1103/PhysRevLett.126.251601",
    journal = "Phys. Rev. Lett.",
    volume = "126",
    number = "25",
    pages = "251601",
    year = "2021"
}

@article{Csaki:2022cyg,
    author = "Cs\'aki, Csaba and Gomes, Andrew and Murayama, Hitoshi and Noether, Bea and Varier, Digvijay Roy and Telem, Ofri",
    title = "{Guide to anomaly-mediated supersymmetry-breaking QCD}",
    eprint = "2212.03260",
    archivePrefix = "arXiv",
    primaryClass = "hep-th",
    doi = "10.1103/PhysRevD.107.054015",
    journal = "Phys. Rev. D",
    volume = "107",
    number = "5",
    pages = "054015",
    year = "2023"
}

@article{Jarlskog:1985ht,
    author = "Jarlskog, C.",
    title = "{Commutator of the Quark Mass Matrices in the Standard Electroweak Model and a Measure of Maximal CP Nonconservation}",
    reportNumber = "USIP-85-14",
    doi = "10.1103/PhysRevLett.55.1039",
    journal = "Phys. Rev. Lett.",
    volume = "55",
    pages = "1039",
    year = "1985"
}

@article{Valenti:2021xjp,
    author = "Valenti, Alessandro and Vecchi, Luca",
    title = "{Super-soft CP violation}",
    eprint = "2106.09108",
    archivePrefix = "arXiv",
    primaryClass = "hep-ph",
    doi = "10.1007/JHEP07(2021)152",
    journal = "JHEP",
    volume = "07",
    number = "152",
    pages = "152",
    year = "2021"
}

@article{Haber:2012np,
    author = "Haber, Howard E. and Surujon, Ze'ev",
    title = "{A Group-theoretic Condition for Spontaneous CP Violation}",
    eprint = "1201.1730",
    archivePrefix = "arXiv",
    primaryClass = "hep-ph",
    reportNumber = "SCIPP-12-01, UCI-TR-2012-01",
    doi = "10.1103/PhysRevD.86.075007",
    journal = "Phys. Rev. D",
    volume = "86",
    pages = "075007",
    year = "2012"
}

@article{Strominger:1985it,
    author = "Strominger, A. and Witten, Edward",
    title = "{New Manifolds for Superstring Compactification}",
    reportNumber = "PRINT-85-0452 (IAS,PRINCETON)",
    doi = "10.1007/BF01216094",
    journal = "Commun. Math. Phys.",
    volume = "101",
    pages = "341",
    year = "1985"
}

@article{Dine:1992ya,
    author = "Dine, Michael and Leigh, Robert G. and MacIntire, Douglas A.",
    title = "{Of CP and other gauge symmetries in string theory}",
    eprint = "hep-th/9205011",
    archivePrefix = "arXiv",
    reportNumber = "SCIPP-92-16",
    doi = "10.1103/PhysRevLett.69.2030",
    journal = "Phys. Rev. Lett.",
    volume = "69",
    pages = "2030--2032",
    year = "1992"
}

@article{Choi:1992xp,
    author = "Choi, Ki-woon and Kaplan, David B. and Nelson, Ann E.",
    title = "{Is CP a gauge symmetry?}",
    eprint = "hep-ph/9205202",
    archivePrefix = "arXiv",
    reportNumber = "UCSD-PTH-92-11",
    doi = "10.1016/0550-3213(93)90082-Z",
    journal = "Nucl. Phys. B",
    volume = "391",
    pages = "515--530",
    year = "1993"
}

@article{Chakdar:2013tca,
    author = "Chakdar, Shreyashi and Ghosh, Kirtiman and Nandi, S. and Rai, Santosh Kumar",
    title = "{Collider signatures of mirror fermions in the framework of a left-right mirror model}",
    eprint = "1305.2641",
    archivePrefix = "arXiv",
    primaryClass = "hep-ph",
    doi = "10.1103/PhysRevD.88.095005",
    journal = "Phys. Rev. D",
    volume = "88",
    number = "9",
    pages = "095005",
    year = "2013"
}

@article{Hall:2018let,
    author = "Hall, Lawrence J. and Harigaya, Keisuke",
    title = "{Implications of Higgs Discovery for the Strong CP Problem and Unification}",
    eprint = "1803.08119",
    archivePrefix = "arXiv",
    primaryClass = "hep-ph",
    doi = "10.1007/JHEP10(2018)130",
    journal = "JHEP",
    volume = "10",
    pages = "130",
    year = "2018"
}

@article{Dunsky:2019api,
    author = "Dunsky, David and Hall, Lawrence J. and Harigaya, Keisuke",
    title = "{Higgs Parity, Strong CP, and Dark Matter}",
    eprint = "1902.07726",
    archivePrefix = "arXiv",
    primaryClass = "hep-ph",
    doi = "10.1007/JHEP07(2019)016",
    journal = "JHEP",
    volume = "07",
    pages = "016",
    year = "2019"
}

@article{Craig:2020bnv,
    author = "Craig, Nathaniel and Garcia Garcia, Isabel and Koszegi, Giacomo and McCune, Amara",
    title = "{P not PQ}",
    eprint = "2012.13416",
    archivePrefix = "arXiv",
    primaryClass = "hep-ph",
    doi = "10.1007/JHEP09(2021)130",
    journal = "JHEP",
    volume = "09",
    pages = "130",
    year = "2021"
}

@article{deVries:2021pzl,
    author = "de Vries, Jordy and Draper, Patrick and Patel, Hiren H.",
    title = "{Do Minimal Parity Solutions to the Strong $CP$ Problem Work?}",
    eprint = "2109.01630",
    archivePrefix = "arXiv",
    primaryClass = "hep-ph",
    month = "9",
    year = "2021"
}

@article{ATLAS:2022hnn,
    author = "Aad, Georges and others",
    collaboration = "ATLAS",
    title = "{Search for pair-production of vector-like quarks in pp collision events at s=13 TeV with at least one leptonically decaying Z boson and a third-generation quark with the ATLAS detector}",
    eprint = "2210.15413",
    archivePrefix = "arXiv",
    primaryClass = "hep-ex",
    reportNumber = "CERN-EP-2021-207",
    doi = "10.1016/j.physletb.2023.138019",
    journal = "Phys. Lett. B",
    volume = "843",
    pages = "138019",
    year = "2023"
}

@article{ATLAS:2022tla,
    author = "Aad, Georges and others",
    collaboration = "ATLAS",
    title = "{Search for pair-produced vector-like top and bottom partners in events with large missing transverse momentum in pp collisions with the ATLAS detector}",
    eprint = "2212.05263",
    archivePrefix = "arXiv",
    primaryClass = "hep-ex",
    reportNumber = "CERN-EP-2022-201",
    doi = "10.1140/epjc/s10052-023-11790-7",
    journal = "Eur. Phys. J. C",
    volume = "83",
    number = "8",
    pages = "719",
    year = "2023"
}

@article{Ellis:1978hq,
    author = "Ellis, John R. and Gaillard, Mary K.",
    title = "{Strong and Weak CP Violation}",
    reportNumber = "FERMILAB-PUB-78-066-T",
    doi = "10.1016/0550-3213(79)90297-9",
    journal = "Nucl. Phys. B",
    volume = "150",
    pages = "141--162",
    year = "1979"
}

@article{Khriplovich:1985jr,
    author = "Khriplovich, I. B.",
    title = "{Quark Electric Dipole Moment and Induced $\theta$ Term in the {Kobayashi-Maskawa} Model}",
    reportNumber = "IYF-86-25",
    doi = "10.1016/0370-2693(86)90245-5",
    journal = "Phys. Lett. B",
    volume = "173",
    pages = "193--196",
    year = "1986"
}

@article{Fukugita:1986hr,
    author = "Fukugita, M. and Yanagida, T.",
    title = "{Baryogenesis Without Grand Unification}",
    reportNumber = "RIFP-641",
    doi = "10.1016/0370-2693(86)91126-3",
    journal = "Phys. Lett. B",
    volume = "174",
    pages = "45--47",
    year = "1986"
}

@article{Farina:2016ndq,
    author = "Farina, Marco and Monteux, Angelo and Shin, Chang Sub",
    title = "{Twin mechanism for baryon and dark matter asymmetries}",
    eprint = "1604.08211",
    archivePrefix = "arXiv",
    primaryClass = "hep-ph",
    doi = "10.1103/PhysRevD.94.035017",
    journal = "Phys. Rev. D",
    volume = "94",
    number = "3",
    pages = "035017",
    year = "2016"
}

@article{Feng:2020urb,
    author = "Feng, Wan-Zhe and Yu, Jiang-Hao",
    title = "{Twin cogenesis}",
    eprint = "2005.06471",
    archivePrefix = "arXiv",
    primaryClass = "hep-ph",
    doi = "10.1088/1572-9494/acbb5b",
    journal = "Commun. Theor. Phys.",
    volume = "75",
    number = "4",
    pages = "045201",
    year = "2023"
}

@article{Kilic:2021zqu,
    author = "Kilic, Can and Verhaaren, Christopher B. and Youn, Taewook",
    title = "{Twin quark dark matter from cogenesis}",
    eprint = "2109.03248",
    archivePrefix = "arXiv",
    primaryClass = "hep-ph",
    reportNumber = "UTTG 16-2021",
    doi = "10.1103/PhysRevD.104.116018",
    journal = "Phys. Rev. D",
    volume = "104",
    number = "11",
    pages = "116018",
    year = "2021"
}

@article{Csaki:2023yas,
    author = "Cs\'aki, Csaba and Tito D'Agnolo, Raffaele and Gupta, Rick S. and Kuflik, Eric and Roy, Tuhin S. and Ruhdorfer, Maximilian",
    title = "{On the dynamical origin of the $\eta'$ potential and the axion mass}",
    eprint = "2307.04809",
    archivePrefix = "arXiv",
    primaryClass = "hep-ph",
    doi = "10.1007/JHEP10(2023)139",
    journal = "JHEP",
    volume = "10",
    pages = "139",
    year = "2023"
}

@article{Csaki:2025atr,
    author = "Cs\'aki, Csaba and Roy, Tuhin S. and Ruhdorfer, Maximilian and Youn, Taewook",
    title = "{Dynamical Up-quark Mass Generation in QCD-like theories}",
    eprint = "2505.07953",
    archivePrefix = "arXiv",
    primaryClass = "hep-ph",
    month = "5",
    year = "2025"
}

@article{Dine:1993qm,
    author = "Dine, Michael and Leigh, Robert G. and Kagan, Alex",
    title = "{Supersymmetry and the Nelson-Barr mechanism}",
    eprint = "hep-ph/9303296",
    archivePrefix = "arXiv",
    reportNumber = "SLAC-PUB-6090, SCIPP-93-05",
    doi = "10.1103/PhysRevD.48.2214",
    journal = "Phys. Rev. D",
    volume = "48",
    pages = "2214--2223",
    year = "1993"
}

@article{Perez:2023zin,
    author = "Perez, Pavel Fileviez and Murgui, Clara and Wise, Mark B.",
    title = "{Automatic Nelson-Barr solutions to the strong CP puzzle}",
    eprint = "2302.06620",
    archivePrefix = "arXiv",
    primaryClass = "hep-ph",
    reportNumber = "CALT-TH/2023-004",
    doi = "10.1103/PhysRevD.108.015010",
    journal = "Phys. Rev. D",
    volume = "108",
    number = "1",
    pages = "015010",
    year = "2023"
}

@article{Murgui:2025scx,
    author = "Murgui, Clara and Patrone, Samuel",
    title = "{Leptogenesis in automatic Nelson-Barr models}",
    eprint = "2506.18963",
    archivePrefix = "arXiv",
    primaryClass = "hep-ph",
    reportNumber = "CALT-TH/2025-020, CERN-TH-2025-118",
    month = "6",
    year = "2025"
}

@article{Hall:2024qqe,
    author = "Hall, Lawrence and Manzari, Claudio Andrea and Noether, Bea",
    title = "{Strong CP and flavor in multi-Higgs theories}",
    eprint = "2407.14585",
    archivePrefix = "arXiv",
    primaryClass = "hep-ph",
    doi = "10.1103/rxml-5gd2",
    journal = "Phys. Rev. D",
    volume = "111",
    number = "11",
    pages = "115012",
    year = "2025"
}

@article{Bonnefoy:2025rvo,
    author = "Bonnefoy, Quentin and Hall, Lawrence J. and Manzari, Claudio Andrea and Noether, Bea",
    title = "{Two Higgs Doublet Solutions to the Strong CP Problem}",
    eprint = "2506.13853",
    archivePrefix = "arXiv",
    primaryClass = "hep-ph",
    month = "6",
    year = "2025"
}

@article{Valenti:2021rdu,
    author = "Valenti, Alessandro and Vecchi, Luca",
    title = "{The CKM phase and $ \overline{\theta} $ in Nelson-Barr models}",
    eprint = "2105.09122",
    archivePrefix = "arXiv",
    primaryClass = "hep-ph",
    doi = "10.1007/JHEP07(2021)203",
    journal = "JHEP",
    volume = "07",
    number = "203",
    pages = "203",
    year = "2021"
}

@article{Vecchi:2025qie,
    author = "Vecchi, Luca",
    title = "{When CP requires $\bar{\theta}=0$, not $\bar{\theta}=\pi$}",
    eprint = "2507.10680",
    archivePrefix = "arXiv",
    primaryClass = "hep-ph",
    month = "7",
    year = "2025"
}

@article{Lillard:2018fdt,
    author = "Lillard, Benjamin and Tait, Tim M. P.",
    title = "{A High Quality Composite Axion}",
    eprint = "1811.03089",
    archivePrefix = "arXiv",
    primaryClass = "hep-ph",
    reportNumber = "UCI-HEP-TR-2018-12",
    doi = "10.1007/JHEP11(2018)199",
    journal = "JHEP",
    volume = "11",
    pages = "199",
    year = "2018"
}

@article{Randall:1992ut,
    author = "Randall, Lisa",
    title = "{Composite axion models and Planck scale physics}",
    reportNumber = "MIT-CTP-2074",
    doi = "10.1016/0370-2693(92)91928-3",
    journal = "Phys. Lett. B",
    volume = "284",
    pages = "77--80",
    year = "1992"
}

@article{DiLuzio:2017tjx,
    author = "Di Luzio, Luca and Nardi, Enrico and Ubaldi, Lorenzo",
    title = "{Accidental Peccei-Quinn symmetry protected to arbitrary order}",
    eprint = "1704.01122",
    archivePrefix = "arXiv",
    primaryClass = "hep-ph",
    reportNumber = "IPPP-17-29",
    doi = "10.1103/PhysRevLett.119.011801",
    journal = "Phys. Rev. Lett.",
    volume = "119",
    number = "1",
    pages = "011801",
    year = "2017"
}

@article{Lillard:2017cwx,
    author = "Lillard, Benjamin and Tait, Tim M. P.",
    title = "{A Composite Axion from a Supersymmetric Product Group}",
    eprint = "1707.04261",
    archivePrefix = "arXiv",
    primaryClass = "hep-ph",
    reportNumber = "UCI-HEP-TR-2017-07",
    doi = "10.1007/JHEP11(2017)005",
    journal = "JHEP",
    volume = "11",
    pages = "005",
    year = "2017"
}

@article{Contino:2021ayn,
    author = "Contino, Roberto and Podo, Alessandro and Revello, Filippo",
    title = "{Chiral models of composite axions and accidental Peccei-Quinn symmetry}",
    eprint = "2112.09635",
    archivePrefix = "arXiv",
    primaryClass = "hep-ph",
    doi = "10.1007/JHEP04(2022)180",
    journal = "JHEP",
    volume = "04",
    pages = "180",
    year = "2022"
}

@article{Cheng:2001ys,
    author = "Cheng, Hsin-Chia and Kaplan, David Elazzar",
    title = "{Axions and a gauged Peccei-Quinn symmetry}",
    eprint = "hep-ph/0103346",
    archivePrefix = "arXiv",
    reportNumber = "EFI-2001-09, ANL-HEP-PR-01-021",
    month = "3",
    year = "2001"
}

@article{Kim:1984pt,
    author = "Kim, Jihn E.",
    title = "{A Composite Invisible Axion}",
    reportNumber = "SNUHE-84-02",
    doi = "10.1103/PhysRevD.31.1733",
    journal = "Phys. Rev. D",
    volume = "31",
    pages = "1733",
    year = "1985"
}

@article{Kaplan:1985dv,
    author = "Kaplan, David B.",
    title = "{Opening the Axion Window}",
    reportNumber = "HUTP-85/A014",
    doi = "10.1016/0550-3213(85)90319-0",
    journal = "Nucl. Phys. B",
    volume = "260",
    pages = "215--226",
    year = "1985"
}

@article{Choi:1985cb,
    author = "Choi, Kiwoon and Kim, Jihn E.",
    title = "{Dynamical Axion}",
    reportNumber = "SNUHE-84-05-REV, SNUHE-84-05",
    doi = "10.1103/PhysRevD.32.1828",
    journal = "Phys. Rev. D",
    volume = "32",
    pages = "1828",
    year = "1985"
}

@article{Fukuda:2017ylt,
    author = "Fukuda, Hajime and Ibe, Masahiro and Suzuki, Motoo and Yanagida, Tsutomu T.",
    title = "{A ''gauged'' $U(1)$ Peccei{\textendash}Quinn symmetry}",
    eprint = "1703.01112",
    archivePrefix = "arXiv",
    primaryClass = "hep-ph",
    reportNumber = "IPMU-17-0040",
    doi = "10.1016/j.physletb.2017.05.071",
    journal = "Phys. Lett. B",
    volume = "771",
    pages = "327--331",
    year = "2017"
}

@article{Kuchimanchi:2025pqb,
    author = "Kuchimanchi, Ravi",
    title = "{Parity solves the Strong CP problem}",
    eprint = "2506.01911",
    archivePrefix = "arXiv",
    primaryClass = "hep-ph",
    month = "6",
    year = "2025"
}

@article{Rosenzweig:1979ay,
    author = "Rosenzweig, C. and Schechter, J. and Trahern, C. G.",
    editor = "Brezin, E. and Wadia, S. R.",
    title = "{Is the Effective Lagrangian for QCD a Sigma Model?}",
    reportNumber = "SU-4217-148, COO-3533-148",
    doi = "10.1103/PhysRevD.21.3388",
    journal = "Phys. Rev. D",
    volume = "21",
    pages = "3388",
    year = "1980"
}

@article{DiVecchia:1980yfw,
    author = "Di Vecchia, P. and Veneziano, G.",
    title = "{Chiral Dynamics in the Large n Limit}",
    reportNumber = "CERN-TH-2814",
    doi = "10.1016/0550-3213(80)90370-3",
    journal = "Nucl. Phys. B",
    volume = "171",
    pages = "253--272",
    year = "1980"
}

@article{Nath:1979ik,
    author = "Nath, Pran and Arnowitt, Richard L.",
    title = "{The U(1) Problem: Current Algebra and the Theta Vacuum}",
    reportNumber = "NUB-2417, HUTP-79/A084",
    doi = "10.1103/PhysRevD.23.473",
    journal = "Phys. Rev. D",
    volume = "23",
    pages = "473",
    year = "1981"
}

@article{Witten:1980sp,
    author = "Witten, Edward",
    title = "{Large N Chiral Dynamics}",
    reportNumber = "HUTP-80/A005",
    doi = "10.1016/0003-4916(80)90325-5",
    journal = "Annals Phys.",
    volume = "128",
    pages = "363",
    year = "1980"
}

@article{Holdom:1984sk,
    author = "Holdom, Bob",
    title = "{Techniodor}",
    reportNumber = "NSF-ITP-84-148",
    doi = "10.1016/0370-2693(85)91015-9",
    journal = "Phys. Lett. B",
    volume = "150",
    pages = "301--305",
    year = "1985"
}

@article{Yamawaki:1985zg,
    author = "Yamawaki, Koichi and Bando, Masako and Matumoto, Ken-iti",
    title = "{Scale Invariant Technicolor Model and a Technidilaton}",
    reportNumber = "DPNU-85-47",
    doi = "10.1103/PhysRevLett.56.1335",
    journal = "Phys. Rev. Lett.",
    volume = "56",
    pages = "1335",
    year = "1986"
}

@article{Appelquist:1986an,
    author = "Appelquist, Thomas W. and Karabali, Dimitra and Wijewardhana, L. C. R.",
    title = "{Chiral Hierarchies and the Flavor Changing Neutral Current Problem in Technicolor}",
    reportNumber = "YTP-86/11",
    doi = "10.1103/PhysRevLett.57.957",
    journal = "Phys. Rev. Lett.",
    volume = "57",
    pages = "957",
    year = "1986"
}
\bibliographystyle{utphys}
}

\end{document}